\documentclass[12pt]{article}
\usepackage{latexsym}
\usepackage{amsmath,amsfonts}
\usepackage{times}
\allowdisplaybreaks[4]
\catcode`\@=11

\newcount\hour
\newcount\minute
\newtoks\amorpm \hour=\time\divide\hour by 60\minute
=\time{\multiply\hour by 60 \global\advance\minute by-\hour}
\edef\standardtime{{\ifnum\hour<12 \global\amorpm={am}%
        \else\global\amorpm={pm}\advance\hour by-12 \fi
        \ifnum\hour=0 \hour=12 \fi
        \number\hour:\ifnum\minute<10
        0\fi\number\minute\the\amorpm}}
\edef\militarytime{\number\hour:\ifnum\minute<10
0\fi\number\minute}

\def\draftlabel#1{{\@bsphack\if@filesw {\let\thepage\relax
   \xdef\@gtempa{\write\@auxout{\string
      \newlabel{#1}{{\@currentlabel}{\thepage}}}}}\@gtempa
   \if@nobreak \ifvmode\nobreak\fi\fi\fi\@esphack}
        \gdef\@eqnlabel{#1}}
\def\@eqnlabel{}
\def\@vacuum{}
\def\marginnote#1{}
\def\draftmarginnote#1{\marginpar{\raggedright\scriptsize\tt#1}}
\def\draft{
        \pagestyle{plain}
        \overfullrule=2pt
        \oddsidemargin -.5truein
        \def\@oddhead{\sl \phantom{\today\quad\militarytime} \hfil
        \smash{\Large\sl DRAFT} \hfil \today\quad\militarytime}
        \let\@evenhead\@oddhead
        \let\label=\draftlabel
        \let\marginnote=\draftmarginnote
        \def\ps@empty{\let\@mkboth\@gobbletwo
        \def\@oddfoot{\hfil \smash{\Large\sl DRAFT} \hfil}
        \let\@evenfoot\@oddhead}
        \def\@eqnnum{(\theequation)\rlap{\kern\marginparsep\tt\@eqnlabel}%
        \global\let\@eqnlabel\@vacuum}  }

\newcommand{\rf}[1]{(\ref{#1})}
\renewcommand{\theequation}{\thesection.\arabic{equation}}
\renewcommand{\thefootnote}{\fnsymbol{footnote}}
\newcommand{\newsection}{    % Numeration of eqs. is automatic
\setcounter{equation}{0}\section}

\def\appendix#1{\addtocounter{section}{1}\setcounter{equation}{0}
\renewcommand{\thesection}{\Alph{section}}
\section*{Appendix \thesection\protect\indent \parbox[t]{11.15cm}{#1}}
\addcontentsline{toc}{section}{Appendix \thesection\ \ \ #1}}

\def\nline{\,\nabla\kern -0.7em\raise0.2ex\hbox{/}\,\,}
\def\yline{\,y\kern -0.47em /}
\def\aline{\,a\kern -0.49em /}
\def\parline{\,\partial\kern -0.55em /\,\,}

\newcommand{\Eo}{\mathbb{E}}

\newcommand{\No}{\mathbb{N}}
\newcommand{\Po}{\mathbb{P}}
\newcommand{\So}{\mathbb{S}}

\newcommand{\Zo}{\mathbb{Z}}

\def\be{\begin{equation}}
\def\ee{\end{equation}}
\def\beq{\begin{eqnarray}}
\def\eeq{\end{eqnarray}}

\def\Bsm{{\scriptscriptstyle B}}
\def\FBsm{{\scriptscriptstyle FB}}
\def\Rsm{{\scriptscriptstyle R}}
\def\Lsm{{\scriptscriptstyle L}}
\def\HDsm{{\scriptscriptstyle HD}}

\def\smone{{\scriptscriptstyle (1)}}
\def\smtwo{{\scriptscriptstyle (2)}}

\def\smpt{{\scriptscriptstyle [2]}}
\def\smpth{{\scriptscriptstyle [3]}}

\def\smpn{{\scriptscriptstyle [n]}}

\def\Jbf{{\bf J}}

\def\Pbf{{\bf P}}
\def\Sbf{{\bf S}}

\def\ibf{{\bf i}}
\def\iibf{{\bf ii}}
\def\iiibf{{\bf iii}}
\def\ivbf{{\bf iv}}
\def\vbf{{\bf v}}
\def\vibf{{\bf vi}}
\def\viibf{{\bf vii}}

\def\MM{{\cal M}}
\def\NN{{\cal N}}
\def\PP{{\cal P}}

\def\half{\frac{1}{2}}

\def\Cb{{\bar{C}}}
\def\Lb{{\bar{L}}}

\def\Vb{{\bar{V}}}

\def\irm{{\rm i}}

\def\f{{\rm f}}

\def\dyn{{\rm dyn}}

\def\diff{{\rm diff}}

\def\crit{{\rm crit}}

\def\pvec{\vec{p}}

\def\betach{\check{\beta}}

\begin{document}

%\draft

\begin{flushright}
FIAN-TD-2020-15  \ \ \ \ \ \ \\
arXiv: 2005.12242 V2
\end{flushright}

\vspace{1cm}

\begin{center}

{\Large \bf Cubic interactions of arbitrary spin fields

\medskip
in 3d flat space}

\vspace{2.5cm}

R.R. Metsaev\footnote{ E-mail: metsaev@lpi.ru }

\vspace{1cm}

{\it Department of Theoretical Physics, P.N. Lebedev Physical
Institute, \\ Leninsky prospect 53,  Moscow 119991, Russia }

\vspace{3cm}

{\bf Abstract}

\end{center}

Using light-cone gauge formulation, massive arbitrary spin irreducible fields and massless (scalar and one-half spin) fields in three-dimensional flat space are considered. Both the integer spin and half-integer spin fields are studied.  For such fields, we provide classification for cubic interactions and obtain explicit expressions for all cubic interaction vertices. We study two forms of the cubic interaction vertices which we refer to as first-derivative form and higher-derivative form. All cubic interaction vertices are built by using the first-derivative form.

\vspace{3cm}


\newpage
\renewcommand{\thefootnote}{\arabic{footnote}}
\setcounter{footnote}{0}

%%%%%%%%%%%%%%%%%%%%%%%%%%%%%%%%%%%%%%%%%%%
\section{ \large Introduction}
%%%%%%%%%%%%%%%%%%%%%%%%%%%%%%

Unitary arbitrary integer and half-integer spin irreducible representations of Poincar\'e algebra in three dimensions are associated with the respective bosonic and fermionic massive fields propagating in the $R^{2,1}$ space. Lagrangian description of free bosonic and fermionic arbitrary spin massive irreducible fields propagating in the $R^{2,1}$ space was obtained long ago in Ref.\cite{Tyutin:1997yn}.
We recall that, in three dimensions, massless fields with spin equal or greater than $\frac{3}{2}$ do not propagate and these fields are not associated with unitary representations of the Poincar\'e algebra $iso(2,1)$. Namely, for massless fields in three dimensions, only scalar, vector and spin one-half fields propagate and only these fields are associated with unitary irreps of the Poincar\'e algebra $iso(2,1)$.
In this paper, we are interested in cubic interactions only for those fields in $R^{2,1}$ which are associated with unitary irreps of the Poincar\'e algebra $iso(2,1)$. Also, we racall, that, in light-cone gauge approach, a massless vector field in $3d$ is treated as a massless scalar field.
In view of above-said, we deal with arbitrary integer and half-integer spin massive fields and low spin (scalar and spin one-half) massless fields.%
\footnote{ For study of various aspects of massless higher-spin dynamics in $R^{2,1}$ and $AdS_3$ spaces, the reader can consult the (incomplete) list of Refs.\cite{Vasiliev:1995dn}-\cite{Mkrtchyan:2017ixk}.
}
Our aim is to construct all cubic interaction vertices for such fields.
In Refs.\cite{Metsaev:2005ar,Metsaev:2007rn}, for light-cone gauge bosonic and fermionic arbitrary spin massive and massless fields propagating in $R^{d-1,1}$, we studied cubic interactions in flat space for the case of $d\geq 4$.%
\footnote{ Lorentz covariant formulation for all light-cone gauge cubic vertices of massless fields in Ref.\cite{Metsaev:2005ar} was obtained in  Refs.\cite{Fotopoulos:2010ay,Metsaev:2012uy}. BRST-BV formulation for all light-cone gauge cubic vertices of massive and massless fields in Ref.\cite{Metsaev:2005ar} was obtained in Ref.\cite{Metsaev:2012uy}. Recent discussion of this theme and extensive list of references may be found in Ref.\cite{Khabarov:2020bgr}.
}
This is to say that, in this paper, we are going to extend results in Refs.\cite{Metsaev:2005ar,Metsaev:2007rn} to the case of light-cone gauge bosonic and fermionic arbitrary spin massive fields and low spin (scalar and spin one-half) massless fields propagating in  $R^{2,1}$ space.

Before proceeding to the main theme in this paper we briefly mention our two long term motivations for our study of massive fields in $3d$ space. First, in view of simplicity of cubic vertices for light-cone gauge massive fields in $R^{2,1}$ obtained in this paper, we believe that our results may be helpful in the search of yet unknown interesting models of higher-spin massive fields in $R^{2,1}$ and their counterparts in higher dimensions. Second, one expects that arbitrary spin AdS massive fields and low spin massless AdS field form spectrum of states of AdS superstring.
We think then that our results for cubic vertices of light-cone gauge massive fields in $R^{2,1}$
may serve as a good starting point  for the study of cubic vertices of light-cone gauge massive fields in $AdS_3$ and hence may find applications in study of superstring in $AdS_3$.%
\footnote{ Light-cone gauge superstring action in $AdS_3$ was considered in Ref.\cite{Metsaev:2000mv}. Interesting use of light-cone gauge approach for studying 3-point function of AdS superstring may be found in Ref.\cite{Klose:2011rm}.
}
Other long term motivations for our study in this paper may be found in Conclusions.

This paper is organized as follows.

In Sec.\ref{sec-02}, we review the well known light-cone gauge description of arbitrary spin massive bosonic and fermionic and low spin (scalar and half-integer) massless fields propagating in $R^{2,1}$.
In Sec.\ref{sec-03npoint}, we describe restrictions on $n$-point interaction vertices imposed by kinematical symmetries of the Poincar\'e algebra $iso(2,1)$.
Sec.\ref{sec-04} is devoted to equations for cubic vertices. We start with  the presentation of  restrictions imposed by kinematical and dynamical symmetries of the Poincar\'e algebra on cubic interaction vertices. After that, we discuss light-cone gauge dynamical principle. Finally we present our complete system of equations which allows us to  determine the cubic vertices unambiguously.
In Sec.\ref{meth-sec-05}, we present our method for solving the complete system of equations for cubic vertices.
In Sec.\ref{bos-sec-06}, we consider cubic interactions for bosonic arbitrary spin massive fields and massless scalar fields. Using our method, we present all solutions for complete system  of equations for cubic vertices obtained in Sec.\ref{sec-04}.
Sec.\ref{fer-bos-sec-07} is devoted to fermi-bose interactions for two fermionic and one bosonic fields. For such interactions, we present all solutions for cubic vertices.
In Sec.\ref{fin-concl}, we present our conclusions.
In Appendix A, we outline some technical details of the derivation of cubic vertices.
In Appendix B, we show how some our vertices can be obtained from vertices of massless fields in $R^{3,1}$ by using dimensional reduction. Also a proposal is made for massive higher-spin theories in $R^{2,1}$.

%%%%%%%%%%%%%%%%%%%%%%%%%%%%%%%%%%%%%%%%%%%%%%%%%%%%%%%%%%%%%%%%%%%%%%%%%%%%%%%
%%%%%%%%%%%%%%%%%%%%%%%%%%%%%%%%%%%%%%%%%%%%%%%%%%%%%%%%%%%%%%%%%%%%%%%%%%%%%%%
\newsection{ \large Light-cone gauge formulation of free massive and massless fields in 3d flat space }\label{sec-02}
%%%%%%%%%%%%%%%%%%%%%%%%%%%%%%%%%%%%%%%%%%%%%%%%%%%%%%%%%%%%%%%%%%%%%%%%%%%%%%%
%%%%%%%%%%%%%%%%%%%%%%%%%%%%%%%%%%%%%%%%%%%%%%%%%%%%%%%%%%%%%%%%%%%%%%%%%%%%%%%

\noindent {\bf Poincar\'e algebra $iso(2,1)$ in light-cone frame}. In Ref.\cite{Dirac:1949cp}, it has been noted that the problem of finding a light-cone gauge dynamical system  amounts to a problem of finding a light cone gauge solution for commutators of a space-time symmetry algebra. For theories of fields propagating in the $R^{2,1}$ space, the space-time symmetries are associated with the Poincar\'e algebra $iso(2,1)$. Therefore our aim in this section
is to review a  realization of the Poincar\'e algebra $iso(2,1)$ on a space of light-cone gauge massive and massless fields. We start with the description  of the Poincar\'e algebra $iso(2,1)$ in the light-cone frame.

In the three-dimensional flat space $R^{2,1}$, the Poincar\'e algebra $iso(2,1)$ is spanned by the three translation generators $P^\mu$, the three generators of the $so(2,1)$ Lorentz algebra denoted as $J^{\mu\nu}$. We use the following form of commutators of the Poincar\'e algebra $iso(2,1)$:
\beq
\label{27022020-man-01} && [P^\mu,\,J^{\nu\rho}]=\eta^{\mu\nu} P^\rho - \eta^{\mu\rho} P^\nu\,,
\qquad {} [J^{\mu\nu},\,J^{\rho\sigma}] = \eta^{\nu\rho} J^{\mu\sigma} + 3\hbox{ terms}\,,
\eeq
where $\eta^{\mu\nu}$ stands for the mostly positive Minkowski metric.
Starting with the Lorentz basis coordinates $x^\mu$, $\mu=0,1,2$, we introduce the light-cone basis coordinates $x^\pm$, $x^1$, where the coordinates $x^\pm$ are defined as
\be \label{27022020-man-02}
x^\pm \equiv \frac{1}{\sqrt{2}}(x^2  \pm x^0)\,.
\ee
From now on, the coordinate $x^+$ is considered as an light-cone time. In the frame  of the light-cone coordinates, a $so(2,1)$ Lorentz algebra vector $X^\mu$ is decomposed as $X^+,X^-$, $X^1$. Also we note that a scalar product of two $so(2,1)$ Lorentz algebra vectors $X^\mu$ and $Y^\mu$ can be decomposed as
\be  \label{27022020-man-03}
\eta_{\mu\nu}X^\mu Y^\nu = X^+Y^- + X^-Y^+ + X^1 Y^1\,.
\ee
Relation \rf{27022020-man-03} tells us, that in the frame of the light-cone frame coordinates, non-vanishing components of the flat metric are given by $\eta_{+-} = \eta_{-+}=1$, $\eta_{11} = 1$. Therefore for the covariant and contravariant components of vector $X^\mu$ we get the relations $X^+=X_-$, $X^-=X_+$, $X^1=X_1$.

In light-cone frame, generators of the Poincar\'e algebra $iso(2,1)$ are separated into the following two groups:
\beq
\label{27022020-man-04} &&
P^+,\quad
P^1,\hspace{0.6cm}
J^{+1},\quad
J^{+-},
\hspace{2cm} \hbox{ kinematical generators};
\\
\label{27022020-man-05}  &&
P^-, \quad
J^{-1},
\hspace{4.6cm} \hbox{ dynamical generators}.
\eeq
Note that, in a field-theoretical realization,  the kinematical generators \rf{27022020-man-04} are quadratic in fields for $x^+=0$,
while, the dynamical generators \rf{27022020-man-05} consist of quadratic and higher order terms in fields.%
\footnote{ Namely, with the exception of $J^{+-}$, all generators \rf{27022020-man-04} are quadratic in fields when $x^+ \ne 0 $, while the $J^{+-}$ takes the form $J^{+-} = G_0 + \irm x^+ P^-$, where  $G_0$ is quadratic in fields.}

In order to get commutators of the Poincar\'e algebra in light-cone frame, we use commutators in \rf{27022020-man-01} and the flat metric $\eta^{\mu\nu}$ which has non-vanishing components given by $\eta^{+-}=\eta^{-+}=1$, $\eta^{11}=1$.
Hermitian conjugation rules for the generators are assumed to be as follows
\be \label{27022020-man-06}
P^{\pm \dagger} = P^\pm, \qquad \ \ P^{1\dagger} = P^1, \qquad  J^{+-\dagger} = - J^{+-}, \quad J^{\pm 1\dagger} = -J^{\pm 1}\,.
\ee
We are going to use a field-theoretical realization of generators of the Poincar\'e algebra on space of light-cone gauge fields. Therefore we now proceed with a review of light-cone gauge description of arbitrary spin massive and low spin massless fields.

\noindent {\bf Light-cone gauge massive and massless fields}. To study field theories in three dimensions we use light-cone gauge massive and massless fields which we denote as $\phi_{m,\lambda}(x)$, where an argument $x$ stands for space time-coordinates $x^+$, $x^-$, $x^1$, while
labels $m$ and $\lambda$ denote the respective mass and  spin of the field $\phi_{m,\lambda}(x)$. We note that
\beq
\label{27022020-man-07} &&  \phi_{m,\lambda}(x) \ \hbox{ are bosonic fields for } \
\lambda \in \No_0\,,
\\[-5pt]
\label{27022020-man-08} &&  \phi_{m,\lambda}(x) \ \hbox{ are fermionic fields for } \
\lambda \in \No_0 + \half\,,
\eeq
By definition, all fields given in \rf{27022020-man-07},\rf{27022020-man-08}, are real-valued
\be \label{27022020-man-08-a0}
\phi_{m,\lambda}^\dagger(x) = \phi_{m,\lambda}(x)\,.
\ee
Using a shortcut $(m,\lambda)$ for the fields $\phi_{m,\lambda}$ in \rf{27022020-man-07}, \rf{27022020-man-08}, we restrict our study in this paper to the following cases of $m$ and $\lambda$
\beq
\label{27022020-man-08-a1} && (m,s)\,, \quad m \ne 0\,,  \ s\in \No\,, \ \ \hbox{ spin-$s$ massive bosonic field};
\\
\label{27022020-man-08-a2} && (m,0)\,, \quad  m\ne 0  \hspace{1.9cm} \hbox{ massive scalar bosonic field};
\\
\label{27022020-man-08-a3} && (0,0)\,, \hspace{3.6cm} \hbox{ massless scalar bosonic field};
\\
\label{27022020-man-08-a4} && (m,s+\half)\,,\quad  m \ne 0\,,  \ s\in \No\,, \hspace{0.4cm} \hbox{ spin-$(s+\half)$ massive fermionic field}; \hspace{1cm}
\\
\label{27022020-man-08-a5} && (m,\half)\,, \quad m \ne 0\,, \hspace{2.6cm} \hbox{ massive spin-$\half$ fermionic field};
\\
\label{27022020-man-08-a6} && (0,\half)\,, \hspace{4.5cm} \hbox{ massless spin-$\half$ fermionic field}.
\eeq

For fields in \rf{27022020-man-08-a1},\rf{27022020-man-08-a4},\rf{27022020-man-08-a5}, the mass parameter $m$ is allowed to be positive or negative. Fields in \rf{27022020-man-08-a1},\rf{27022020-man-08-a4} with $m>0$ and $m<0$ are refereed to as self-dual and anti-self-dual massive fields respectively. The self-dual massive field $(|m|,\lambda)$ and anti-self-dual massive field $(-|m|,\lambda)$ are associated with different irreducible irreps of the Poincar\'e algebra $iso(2,1)$.

In place of the fields depending on space-time coordinates \rf{27022020-man-07},\rf{27022020-man-08}, we prefer to deal with the fields obtainable by using the Fourier transform with respect to the spatial coordinates $x^-$, $x^1$,
\be \label{27022020-man-11}
\phi_{m,\lambda}(x) = \int \frac{ d^2\pvec }{ 2\pi } e^{\irm(\beta x^- + p x^1)} \phi_{m,\lambda}(x^+,\pvec)\,, \hspace{1cm} d^2\pvec \equiv dp\,d\beta\,,
\ee
where the argument $\pvec$ in $\phi_{m,\lambda}(x^+,p)$, is used to indicate the momenta $\beta$, $p$. In terms of the field $\phi_{m,\lambda}(x^+,p)$, the hermicity condition shown in \rf{27022020-man-08-a0} takes the following form:
\be \label{27022020-man-12}
\phi_{m,\lambda}^\dagger(\pvec) = \phi_{m,\lambda}(-\pvec)\,,
\ee
Note, that, in \rf{27022020-man-11} and below, dependence of the momentum-space fields on the light-cone time $x^+$ is implicit.

\noindent {\bf Realization of the Poincar\'e algebra on fields}. We now ready to present the well known  field-theoretical realization of the Poincar\'e algebra on the space of massive and massless fields in three dimensions.
This is to say that a realization of the Poincar\'e algebra \rf{27022020-man-01} in terms of differential operators acting on the momentum-space fields $\phi_{m,\lambda}(p)$ \rf{27022020-man-11} is given by the following relations:
\beq
\label{27022020-man-13} && P^1 = p\,,  \qquad   \hspace{0.8cm} P^+=\beta\,,\qquad
P^- = p^-\,, \qquad p^- \equiv - \frac{p^2+m^2}{2\beta}\,,\qquad
\\[-5pt]
\label{27022020-man-14} && J^{+1}= \irm x^+ P^1 + \partial_p\beta\,,  \hspace{2cm} J^{+-} = \irm x^+P^- + \partial_\beta \beta - \half e_\lambda\,,
\\[-5pt]
\label{27022020-man-15} && J^{-1} = -\partial_\beta p + \partial_p p^- + \frac{1}{\beta} M+ \frac{p}{2\beta}e_\lambda\,, \hspace{1cm} M = \irm m \lambda\,,
\\
\label{27022020-man-16} && \hspace{1cm} \partial_\beta\equiv \partial/\partial \beta\,, \quad \partial_p \equiv \partial/\partial p\,,
\\
\label{27022020-man-17} && \hspace{1cm}  e_\lambda =0 \hspace{0.5cm} \hbox{ for integer } \ \lambda\,,\hspace{1.4cm}
e_\lambda = 1 \hspace{0.5cm} \hbox{ for half-integer } \ \lambda\,.
\eeq

We now note that, to quadratic order in the fields $\phi_{m,\lambda}(p)$, the field-theoretical realization of  the Poincar\'e algebra generators \rf{27022020-man-01} takes the form
\be \label{27022020-man-18}
G_\smpt   =  \int d^2 \pvec\,\,   \beta^{e_{\lambda+\half}} \phi_{m,\lambda}^\dagger G_\diff \phi_{m,\lambda}\,,
\ee
where $G_\diff$ stands for the differential operators given in
\rf{27022020-man-13}-\rf{27022020-man-15}, while the notation
$G_\smpt$ \rf{27022020-man-18} is used for the field-theoretical representation for the generators of the Poincar\'e algebra \rf{27022020-man-01}.

By definition, the fields $\phi_{m,\lambda}$ given in \rf{27022020-man-07} and \rf{27022020-man-08} satisfy the respective Poisson-Dirac equal-time commutation and anti-commutation relations
\beq
\label{27022020-man-19} && [\phi_{m,\lambda}(\pvec), \phi_{m,\lambda'}(\pvec\,{}')] = \frac{1}{2\beta}\delta^2(\pvec+\pvec\,{}')  \delta_{\lambda\lambda'}\,,  \hspace{0.9cm} \hbox{ for } \ \lambda \in \No_0\,,
\\
\label{27022020-man-20} && \{\phi_{m,\lambda}(\pvec), \phi_{m,\lambda'}(\pvec\,{}')\} = \half \delta^2(\pvec+\pvec\,{}') \delta_{\lambda\lambda'}\,, \hspace{1cm} \hbox{ for } \ \lambda \in \No_0 + \half\,.
\eeq
Taking into account relations above-given it is easy to check the following standard equal-time commutation relations between the fields and the Poincar\'e algebra generators
\be \label{27022020-man-21}
[ \phi_{m,\lambda},G_\smpt\,] =  G_\diff \phi_{m,\lambda} \,.
\ee
%
%%%%%%%%%%%%%%%%%%%%%%%%%%%%%%%%%%%%%%%%%%%%%%%%%%%%%%%%%%%%%%%%%%%%%%%%%%%%%%%
%%%%%%%%%%%%%%%%%%%%%%%%%%%%%%%%%%%%%%%%%%%%%%%%%%%%%%%%%%%%%%%%%%%%%%%%%%%%%%%
\newsection{ \large General structure of $n$-point vertices} \label{sec-03npoint}
%%%%%%%%%%%%%%%%%%%%%%%%%%%%%%%%%%%%%%%%%%%%%%%%%%%%%%%%%%%%%%%%%%%%%%%%%%%%%%%
%%%%%%%%%%%%%%%%%%%%%%%%%%%%%%%%%%%%%%%%%%%%%%%%%%%%%%%%%%%%%%%%%%%%%%%%%%%%%%%

In this section, we describe restrictions imposed on interacting vertices by the kinematical symmetries of the Poincar\'e algebra $iso(2,1)$.

For interacting fields, dynamical generators the Poincar\'e algebra receive corrections having higher powers of fields. Namely, the dynamical generators $G^\dyn=P^-, J^{-1}$ can be presented as
\be \label{28022020-man-01}
G^\dyn
= \sum_{n=2}^\infty
G_\smpn^\dyn\,,
\ee
where $G_\smpn^\dyn$ stands for a functional
that has $n$ powers of bosonic and fermionic fields.
For $n\geq 3$, we are going to describe restrictions imposed on the dynamical generators $G_\smpn^\dyn$ obtained from commutators between $G_\smpn^\dyn$ and the kinematical generators. Let us discuss the restrictions in turn.

\noindent {\bf Kinematical $P^1$, $P^+$ symmetries.}. Using commutators between the dynamical generators $P^-$, $J^{-1}$ and the kinematical generators $P^1$,  $P^+$,   we find that, for $n\geq 3$,  the dynamical generators $P_\smpn^-$ and $J_\smpn^{-1}$  can be presented as
\beq
\label{28022020-man-02}  && P_\smpn^- = \int\!\! d\Gamma_\smpn\,\,    \Phi_\smpn^*    p_\smpn^- \,,
\\
\label{28022020-man-03}  && J_\smpn^{-1} = \int\!\! d\Gamma_\smpn\,\,  \Phi_\smpn^*   j_\smpn^{-1}  -   \big( \frac{1}{n}\sum_{a=1}^n \partial_{p_a} \Phi_\smpn^* \big)  p_\smpn^-  \,,
\eeq
and we use the following notation
\beq
\label{28022020-man-04} && d\Gamma_\smpn =  (2\pi)^2 \delta^{2}(\sum_{a=1}^n \pvec_a)\prod_{a=1}^n \frac{d^2\pvec_a}{ 2\pi }\,, \qquad d^2 \pvec_a \equiv dp_a d\beta_a\,,
\\
\label{28022020-man-06} && \Phi_\smpn^* =  \prod_{a=1}^n \phi_{m_a,\lambda_a}^\dagger(\pvec_a)\,.
\eeq
Indices $a,b=1,\ldots,n$ label fields entering $n$-point vertex.
The argument $\pvec_a$ in \rf{28022020-man-04},\rf{28022020-man-06} stands for the momenta $p_a$ and $\beta_a$.
Throughout this paper, the density $p_\smpn^-$ is referred to as an $n$-point interaction vertex. For $n=3$, the density $p_\smpth^-$ is referred to as cubic interaction vertex.

\noindent {\bf Kinematical $J^{+-}$-symmetry equations}. Commutators between the generators $P^-$, $J^{-1}$ and the generator $J^{+-}$ lead to following equations for the densities:
\be \label{28022020-man-07}
\sum_{a=1}^n  \big( \beta_a\partial_{\beta_a} +  \half e_{\lambda_a} \big) p_\smpn^- =  0\,,
\hspace{1cm}  \sum_{a=1}^n  \big( \beta_a\partial_{\beta_a} +  \half e_{\lambda_a} \big) j_\smpn^{-1} =  0\,.
\ee

\noindent {\bf Kinematical $J^{+1}$-symmetry equations}. Commutators between the generators $P^-$, $J^{-1}$ and the  generator $J^{+1}$ tell us that the densities $p_\smpn^-$, $j_\smpn^{-1}$ depend on the momenta $p_a$ through new momentum variables $\Po_{ab}$,
\beq
\label{28022020-man-09}
& p_\smpn^- = p_\smpn^-(\Po_{ab}\,, \beta_a)\,,\hspace{1cm} j_\smpn^{-1} = j_\smpn^{-1}(\Po_{ab}\,, \beta_a)\,, &
\\
\label{28022020-man-08} & \Po_{ab}\equiv p_a \beta_b - p_b \beta_a\,. &
\eeq

Relations given in \rf{28022020-man-02}-\rf{28022020-man-09} provide a complete list restrictions imposed by the kinematical symmetries on the $n$-point dynamical generators $P_\smpn^-$, $J_\smpn^{-1}$.
Below we apply the general discussion above-presented for studying cubic vertices which correspond to the case $n=3$.

%%%%%%%%%%%%%%%%%%%%%%%%%%%%%%%%%%%%%%%%%%%%%%%%%%%%%%%%%%%%%%%%%%%%%%%%%%%%%%%
%%%%%%%%%%%%%%%%%%%%%%%%%%%%%%%%%%%%%%%%%%%%%%%%%%%%%%%%%%%%%%%%%%%%%%%%%%%%%%%
\newsection{ \large Complete system of equations for cubic vertices } \label{sec-04}
%%%%%%%%%%%%%%%%%%%%%%%%%%%%%%%%%%%%%%%%%%%%%%%%%%%%%%%%%%%%%%%%%%%%%%%%%%%%%%%
%%%%%%%%%%%%%%%%%%%%%%%%%%%%%%%%%%%%%%%%%%%%%%%%%%%%%%%%%%%%%%%%%%%%%%%%%%%%%%%

We now turn to studying cubic vertices. Our plan in this section is as follows. First, we represent $J^{+-}$ symmetry equations given in \rf{28022020-man-07} in terms of the momenta $\Po_{ab}$ defined in \rf{28022020-man-08}. Second, we consider restrictions imposed by dynamical symmetries. Third, we formulate our dynamical principle which we refer to as light-cone gauge dynamical principle. Finally, we present the complete system equations which allow us to determine the cubic vertices unambiguously.

\noindent {\bf Kinematical symmetries of the cubic densities}. Using the momentum conservation laws
\be  \label{29022020-man-01}
p_1 + p_2 + p_3 = 0\,, \quad \beta_1 +\beta_2 +\beta_3 =0 \,,
\ee
it is easy to check that $\Po_{12}$, $\Po_{23}$, $\Po_{31}$ are expressed in terms of a new momentum $\Po$
\beq
\label{29022020-man-02} & \Po_{12}  =\Po_{23}  = \Po_{31}  = \Po\,,  &
\\
\label{29022020-man-03}  & \Po  \equiv \frac{1}{3}\sum_{a=1,2,3} \betach_a p_a  \,, \qquad
\betach_a\equiv \beta_{a+1}-\beta_{a+2}\,, \quad \beta_a\equiv
\beta_{a+3}\,.&
\eeq
We now see that our cubic densities $p_\smpth^-$, $j_\smpth^{-1}$ are functions of
the momenta $\Po$, $\beta_1$, $\beta_2$, $\beta_3$,
\be  \label{29022020-man-05}
p_\smpth^- = p_\smpth^-(\Po\,, \beta_a)\,, \qquad  j_\smpth^{-1} = j_\smpth^{-1}(\Po\,, \beta_a)\,.
\ee
Thus, the three momenta $p_a$ enter cubic densities \rf{29022020-man-05} through the one momentum $\Po$. We note then that it is this feature of the cubic densities that simplifies the study of cubic interactions.  Let us now represent  equations \rf{28022020-man-07} in terms of the cubic densities given in \rf{29022020-man-05}.

\noindent {\bf $J^{+-}$-symmetry equations}: Taking into account representation for cubic densities in \rf{29022020-man-05}, it is easy to see that, for $n=3$,  equations \rf{28022020-man-07} are represented as
\beq
\label{29022020-man-06} &&  \Jbf^{+-} p_\smpth^- = 0\,,  \hspace{0.7cm} \Jbf^{+-} j_\smpth^{-1} = 0\,,\qquad
\\
\label{29022020-man-07} && \hspace{1cm} \Jbf^{+-} \equiv    \Po \frac{\partial}{\partial \Po} +  \sum_{a=1,2,3}  \big( \beta_a \partial_{\beta_a} + \half e_{\lambda_a}\big)\,.
\eeq

We now turn to studying restrictions imposed by the dynamical symmetries.

\noindent {\bf Dynamical symmetries of the cubic densities}. Restrictions on the interaction vertices obtained  from commutators between the dynamical generators of the Poincar\'e algebra are referred to as dynamical symmetry restrictions in this paper. In our case all that is required is to consider commutator between dynamical generators given in \rf{27022020-man-05}, $[P^-,J^{-1}]=0$. In the cubic approximation, this commutator leads to the relation
\be \label{29022020-man-09}
[P_\smpt^- ,J_\smpth^{-1}] + [P_\smpth^-,J_\smpt^{-1}]=0\,.
\ee
Equation  \rf{29022020-man-09} tells us that the density $j_\smpth^{-1}$ can be expressed in terms of the cubic vertex $p_\smpth^-$,
\beq
\label{29022020-man-10} && j_\smpth^{-1}  = - (\Pbf^-)^{-1} \Jbf^{-1\dagger} p_\smpth^- \,,
\eeq
where operators $\Pbf^-$, $\Jbf^{-1\dagger}$ are defined by the following relations
\beq
\label{29022020-man-11} &&  \Pbf^-  = \frac{\Po^2}{2\beta} - \sum_{a=1,2,3} \frac{m_a^2}{2\beta_a}\,, \hspace{1cm} \beta  \equiv  \beta_1 \beta_2 \beta_3\,,
\\
\label{29022020-man-12}&&    \Jbf^{-1\dagger}  =   - \frac{\Po}{\beta} \No_\beta^E - \MM + \sum_{a=1,2,3}  \frac{\check\beta_a }{6\beta_a} m_a^2 \partial_\Po\,, \qquad \No_\beta^E \equiv \No_\beta + \half \Eo_\lambda\,,
\\
\label{29022020-man-13} && \No_\beta    \equiv  \frac{1}{3}\sum_{a=1,2,3} \check\beta_a \beta_a \partial_{\beta_a}\,, \hspace{1cm}  \Eo_\lambda    \equiv  \frac{1}{3}\sum_{a=1,2,3} \betach_a e_{\lambda_a}\,,  \hspace{1cm}  \MM \equiv  \sum_{a=1,2,3} \frac{1}{\beta_a} M_a\,.
\eeq

Equation \rf{29022020-man-09} exhausts restrictions imposed by the dynamical symmetries of the Poincar\'e algebra on the cubic densities $p_\smpth^-$ and $j_\smpth^{-1}$. Kinematical and dynamical symmetry equations given in \rf{29022020-man-06} and \rf{29022020-man-09} respectively provide complete list of restrictions imposed by Poincar\'e algebra.

\noindent {\bf Light-cone gauge dynamical principle}. Poincar\'e algebra restrictions given in \rf{29022020-man-06}, \rf{29022020-man-10} do not allow us to fix the cubic vertex $p_\smpth^-$ unambiguously. To fix the cubic vertex $p_\smpth^-$ unambiguously we impose additional restrictions on the densities $p_\smpth^-$ and $j_\smpth^{-1}$ which we refer to  as light-cone gauge dynamical principle. We formulate the light-cone gauge dynamical principle as follows:

\noindent \ibf) The cubic densities $p_\smpth^-$, $j_\smpth^{-1}$ should be polynomial in the momentum  $\Po$ and $\beta$-analytic;%
\footnote{ If function $f=f(\beta_1,\beta_2,\beta_3)$ takes the form  $f=g/h$   where $g$, $h$ are polynomials in $\beta_1 ,\beta_2, \beta_3$, then we say that $f$ is $\beta$-analytic. If $f= \sqrt{u} g/h$, where $g$, $h$, $u$ are polynomials in $\beta_1 ,\beta_2, \beta_3$, then we say that $f$ is $\beta$-nonanalytic. Consider
Taylor series expansion for $p_\smpth^- = \sum_{n=0}^K f_n\Po^n$. If all $f_n$ are $\beta$-analytic, then we say that $p_\smpth^-$ is $\beta$-analytic. If some $f_n$ are $\beta$-nonanalytic, then we say that $p_\smpth^-$ is $\beta$-nonanalytic.

}

\noindent \iibf) The cubic vertex $p_\smpth^-$ should satisfy the following restriction
\be \label{29022020-man-14}
p_\smpth^-  \ne  \Pbf^- W\,, \quad W \ \hbox{is polynomial in $\Po$ and $\beta$-analytic},\qquad \qquad
\ee
where $\Pbf^-$ is given in \rf{29022020-man-11}.
Let us explain restriction \rf{29022020-man-14}. Upon field redefinitions, the cubic vertex $p_\smpth^-$ is changed by terms proportional to the quantity $\Pbf^-$ \rf{29022020-man-11} (for discussion, see, e.g., Appendix B in Ref.\cite{Metsaev:2005ar}). Therefore, ignoring restriction in \rf{29022020-man-14} implies that the cubic vertices can be removed by field redefinitions. However our primary interest are the cubic vertices that cannot be removed by field redefinitions. For this reason, we impose the requirement \rf{29022020-man-14}.
We now summarize our discussion equations and restrictions for the cubic densities.

\noindent {\bf Complete system of equations for cubic interaction vertex}. For the cubic vertex given by
\be  \label{29022020-man-15}
p_\smpth^- = p_\smpth^-(\Po\,, \beta_a)
\ee
the complete system of equations is given by
\beq
&& \hspace{-2.5cm} \hbox{ \sf \small Poincar\'e algebra kinematical and dynamical restrictions:}
\nonumber\\
\label{29022020-man-16} && \hspace{-1cm} \Jbf^{+-} p_\smpth^- =0 \,, \hspace{4.7cm} \hbox{kinematical } \  J^{+-}-\hbox{ symmetry};
\\
\label{29022020-man-17} && \hspace{-1cm} j_\smpth^{-1} = - (\Pbf^-)^{-1} \Jbf^{-1\dagger} p_\smpth^- \,, \hspace{2.5cm} \hbox{ dynamical } P^-, J^{-1} \hbox{ symmetries };\qquad
\\
&& \hspace{-2.5cm} \hbox{ \sf\small Light-cone gauge dynamical principle:}
\nonumber\\
\label{29022020-man-18} && \hspace{-1cm} p_\smpth^-\,,    \ j_\smpth^{-1} \hspace{1.2cm} \hbox{ are polynomial in $\Po$ and $\beta$-analytic;}
\\
\label{29022020-man-19} && \hspace{-1cm} p_\smpth^- \ne \Pbf^- W, \hspace{0.5cm} W  \hbox{  is polynomial in $\Po$ and $\beta$-analytic},   \qquad
\eeq
where operator $\Jbf^{+-}$ is given in \rf{29022020-man-07}, while the operators $\Pbf^-$, $\Jbf^{-1\dagger}$ are defined in \rf{29022020-man-11}-\rf{29022020-man-13}

Equations presented in \rf{29022020-man-16}-\rf{29022020-man-19} constitute the complete system of equations which allow us to find all cubic interaction vertices $p_\smpth^-$ and the corresponding densities $j_\smpth^{-1}$ unambiguously.

%%%%%%%%%%%%%%%%%%%%%%%%%%%%%%%%%%%%%%%%%%%%%%%%%%%%%%%%%%%%%%%%%%%%%%%%%%%%%%%%%%%%%%%%%%%
%%%%%%%%%%%%%%%%%%%%%%%%%%%%%%%%%%%%%%%%%%%%%%%%%%%%%%%%%%%%%%%%%%%%%%%%%%%%%%%%%%%%%%%%%%%
\newsection{ \large Method for solving complete system of equations for cubic vertices} \label{meth-sec-05}
%%%%%%%%%%%%%%%%%%%%%%%%%%%%%%%%%%%%%%%%%%%%%%%%%%%%%%%%%%%%%%%%%%%%%%%%%%%%%%%%%%%%%%%%%%
%%%%%%%%%%%%%%%%%%%%%%%%%%%%%%%%%%%%%%%%%%%%%%%%%%%%%%%%%%%%%%%%%%%%%%%%%%%%%%%%%%%%%%%%%%

The most difficult point in the analysis of the equations \rf{29022020-man-16}-\rf{29022020-man-19} is related to the fact that, in general, the cubic vertex is some complicated polynomial in the momentum $\Po$. We note however, {\it in light-cone gauge approach in three dimensions, by using field redefinitions, any cubic vertex can be cast into a polynomial of degree-1 in $\Po$} (see below). Representation for cubic vertex in terms of degree-1 polynomials in $\Po$ will be referred to as first-derivative form of cubic vertex. Thus our first-derivative vertices are degree-1 polynomials in $\Po$. In this paper, we find first-derivative form for all cubic vertices. Besides this, for some wide class of first-derivative cubic vertices, we discuss the procedures which allow us, by using field redefinitions, to generate other particular representation for cubic vertices which we refer to as higher-derivative form of cubic vertices.%
\footnote{ We expect that higher-derivative form of cubic vertices is more convenient as starting point for translation of our light-cone gauge cubic vertices into Lorentz covariant cubic vertices. This is long-term motivation of our interest in studying higher-derivative form of cubic vertices.}
We start with discussion of first-derivative form for the cubic vertex.

\noindent {\bf First-derivative form of cubic vertex}. We find that {\it any cubic vertex} $p_\smpth^-$, which involves at least one massive field, can be presented as
\beq
\label{01032020-man-x01} && p_\smpth^- = \frac{1}{2} (1 + \frac{\irm \Po}{\kappa}) V + \frac{1}{2} (1 - \frac{\irm \Po}{\kappa} ) \Vb\,,
\\
\label{01032020-man-x02} && \hspace{1cm} \kappa^2  \equiv  -\beta \sum_{a=1,2,3} \frac{m_a^2}{\beta_a}\,,\qquad \beta\equiv\beta_1\beta_2\beta_3\,,
\eeq
where two new vertices $V$, $\Vb$ do not depend on the momentum $\Po$. These two new vertices depend only on the momenta $\beta_1$, $\beta_2$, $\beta_3$ and satisfy the following decoupled (diagonalized) equations
\beq
\label{01032020-man-x03} && \frac{\kappa}{\beta}  \No_\beta^E   V +  \irm \MM  V  = 0\,, \hspace{1cm} \frac{\kappa}{\beta}  \No_\beta^E   \Vb -  \irm \MM  \Vb= 0\,.
\\
\label{01032020-man-x04} && \Jbf^{+-} V = 0\,, \qquad \Jbf^{+-} \Vb = 0\,, \qquad \Jbf^{+-} \equiv \sum_{a=1,2,3} ( \beta_a\partial_{\beta_a} + \half e_{\lambda_a})\,, \qquad
\eeq
where we use notation as in \rf{29022020-man-12},\rf{29022020-man-13}. Thus, the complete system of equations is reduced to analysis of decoupled equations for the vertices $V$, $\Vb$ \rf{01032020-man-x03},\rf{01032020-man-x04}. Expression for $j_\smpth^{-1}$ corresponding to cubic vertex \rf{01032020-man-x01} is given by
\be \label{01032020-man-x05}
j_\smpth^{-1} = 2\irm \No_\beta^E V_1\,, \qquad V_1 \equiv \frac{1}{2\kappa}(V- \Vb)\,. %
\ee
Equations \rf{01032020-man-x03},\rf{01032020-man-x04} fix vertices $V$, $\Vb$ uniquely (up to two coupling constants).  If $\kappa$  \rf{01032020-man-x02}  is $\beta$-analytic, then both vertices $V$, $\Vb$ are also $\beta$-analytic, and, in view of \rf{01032020-man-x01}, we obtain two vertices $p_\smpth^-$ \rf{01032020-man-x01}. If $\kappa$ \rf{01032020-man-x02} is $\beta$-nonanalytic, then both vertices $V$, $\Vb$ are also  $\beta$-nonanalytic. For this case, requiring the $p_\smpth^-$ \rf{01032020-man-x01} to be $\beta$-analytic, we get one restriction on two coupling constants. Solving this restriction, we are left with one $\beta$-analytic vertex $p_\smpth^-$. Thus, {\it if $\kappa$ is $\beta$-analytic, then there are two vertices, while, if $\kappa$ is $\beta$-nonanalytic, then there is one vertex}. We see that number of cubic vertices $p_\smpth^-$ depends only on masses and does not depend on spins.

We now outline our method for derivation of equations given in \rf{01032020-man-x01}-\rf{01032020-man-x05}. Our method can be described as a sequence of the following steps.

\noindent \ibf) Cubic vertex $p_\smpth^-$ is defined by module of field redefinitions. We use field redefinitions to choose the most simple representation for cubic vertex. Upon field redefinitions, the cubic vertex is changed as
\be \label{01032020-man-01}
p_\smpth^- \rightarrow p_\smpth^- + \Pbf^- f\,,
\hspace{1cm} f = f(\Po,\beta_a)\,,
\ee
where $f(\Po,\beta_a)$ is polynomial in $\Po$ and $\beta$-analytic, while $\Pbf^-$ is given in \rf{29022020-man-11}. By definition the cubic vertex $p_\smpth^-$ is a finite-order polynomial in $\Po$ and $\beta$-analytic. Taking into account that $\Pbf^-$ is degree-2 polynomial in $\Po$ \rf{29022020-man-11} it is clear from \rf{01032020-man-01} that, by using field redefinition, we can remove in the cubic vertex $p_\smpth^-$ all terms $\Po^q$ with $q\geq 2$. In other words, by using field redefinitions, the cubic vertex $p_\smpth^-$ can be made to be degree-1 polynomial in $\Po$,
\be \label{01032020-man-03}
p_\smpth^- = \irm \Po V_1 + V_0\, \hspace{1cm}  V_1 = V_1(\beta_a)\,, \qquad V_0= V_0(\beta_a)\,,
\ee
where new vertices $V_1$, $V_0$ do not depend on $\Po$ and should be $\beta$-analytic.
The vertex $p_\smpth^-$ \rf{01032020-man-03} satisfies requirement \rf{29022020-man-19}. Thus, by using field redefinitions, we get the simple representation for the vertex  \rf{01032020-man-03} and respect requirement \rf{29022020-man-19}. Note that the possibility to cast any cubic vertex into the first-derivative form \rf{01032020-man-03} exists only in three dimensions.

\noindent \iibf) Using \rf{01032020-man-03}, we now consider equation \rf{29022020-man-17} and requirement for the density $j_\smpth^{-1}$ to be polynomial in $\Po$.
Using expression for $\Jbf^{-1\dagger}$ \rf{29022020-man-12}, we find the following relation:
\beq
\label{01032020-man-05} \Jbf^{-1\dagger}  p_\smpth^- & = & -2\irm \Pbf^-  \No_\beta^E V_1
\nonumber\\
& + & \irm \sum_{a=1,2,3}\big(- \frac{m_a^2}{\beta_a} \No_\beta^E + \frac{\betach_a m_a^2}{6\beta_a}\big) V_1 - \MM  V_0
\nonumber\\
& - & \Po \Big(\frac{1}{\beta} \No_\beta^E V_0 + \irm \MM V_1\Big)\,.
\eeq
Using \rf{01032020-man-05}, we see that equation \rf{29022020-man-17} and requirement for the density $j_\smpth^{-1}$ to be polynomial in $\Po$ amount to the following two equations
\beq
\label{01032020-man-06} && \irm \sum_{a=1,2,3}\big(- \frac{m_a^2}{\beta_a} \No_\beta^E + \frac{\betach_a m_a^2}{6\beta_a}\big) V_1 - \MM  V_0 = 0 \,,
\\
\label{01032020-man-07} && \frac{1}{\beta} \No_\beta^E V_0 + \irm \MM V_1 =0 \,,
\eeq
and the following representation for the density $j_\smpth^{-1}$:
\be \label{01032020-man-08}
j_\smpth^{-1} = 2\irm \No_\beta^E V_1\,.
\ee

\noindent \iiibf) Plugging \rf{01032020-man-03} into \rf{29022020-man-16} we find the following equations for the vertices $V_1$, $V_0$:
\be \label{01032020-man-09}
\Jbf^{+-} V_0 = 0\,, \qquad \big( \Jbf^{+-} +1)V_1 = 0\,, \qquad \Jbf^{+-} \equiv \sum_{a=1,2,3} ( \beta_a\partial_{\beta_a} + \half e_{\lambda_a})\,. \qquad
\ee
Thus, by using cubic vertex \rf{01032020-man-03}, we reduced our complete system of equations \rf{29022020-man-16}-\rf{29022020-man-19} to equations for vertices $V_1$, $V_0$ in \rf{01032020-man-06},\rf{01032020-man-07} and \rf{01032020-man-09}. Equations \rf{01032020-man-09} are simple homogeneity equations. It is the equations \rf{01032020-man-06},\rf{01032020-man-07} that turn out to be complicated for the analysis.

\noindent \ivbf) Vertices $V_1$, $V_0$ satisfy coupled equations \rf{01032020-man-06},\rf{01032020-man-07}. Our basic observation is that these equations can be cast into decoupled (diagonalized) form. Namely, in place of vertices $V_1$, $V_0$, we introduce vertices $V$, $\Vb$ defined by the relations
\be \label{01032020-man-11}
V = V_0  + \kappa V_1\,,
\hspace{1cm} \Vb = V_0 - \kappa V_1\,.
\ee
It is the straightforward exercise to show that two equations for the vertices $V_1$, $V_0$ \rf{01032020-man-06},\rf{01032020-man-07} amount to the decoupled (diagonalized) equations for the vertices $V$ , $\Vb$ given in \rf{01032020-man-x03}. By definition, the vertices $V_1$, $V_0$ are $\beta$-analytic. For arbitrary masses $m_a$, the $\kappa$ \rf{01032020-man-x02} is $\beta$-nonanalytic. Therefore, from \rf{01032020-man-11}, we see that, in general, the vertices $V$ , $\Vb$ are also $\beta$-nonanalytic. This is to say that solution for $V$, $\Vb$ should be chosen so that to get $\beta$-analytic $V_1$, $V_0$.

We note that, for the case of three massless fields, $\kappa=0$. Therefore transformation \rf{01032020-man-11} is not invertible. For analysis of this particular case, we will use basis of vertices $V_0$ and $V_1$.

\noindent {\bf Higher-derivative form of cubic vertex}. We start with the definition of higher-derivative vertices we use in this paper. The first-derivative vertex is degree-1 polynomial in $\Po$. Applying field redefinitions to the first derivative vertex, we obtain a general vertex which is degree-$n$, $n\geq 2$, polynomial in $\Po$. Consider the general vertex for spin $s_1$, $s_2$, $s_3$ bosonic fields and let us introduce quantities $L_a$, $L_{\crit, a}$, $a=1,2,3$, defined below in \rf{01032020-man-15},\rf{02032020-man-24-c3}.
If the general vertex depends on $\Po$ through expressions $L_a^{s_a}$, $L_{\crit, a}^{\epsilon_a s_a}$, $\epsilon_a^2=1$, then we refer to such general vertex as higher-derivative vertex.%
\footnote{ We use the expressions $L_a^{s_a}$ because we expect interrelations between such expressions with linearized curvatures in Lorentz covariant formulations.  We use the expressions $L_{\crit a}^{\epsilon_a s_a}$ because such expressions gives rise upon compactification from $4d$ massless field to $3d$ massive fields (see Appendix B).
}
Fermi-bose higher-derivative vertices involve some additional factor which depends linearly on $\Po$ (see factor $K$ in \rf{01032020-man-14}). The first-derivative vertices do not impose any constraints on spins, while, as we demonstrate below, some higher-derivative vertices impose certain constraints on spins. This is to say that, first-derivative vertices provide us the full list of vertices, while higher-derivative vertices we find in this paper provide us the particular list of vertices.

Some first-derivative vertices can be used to generate their higher-derivative counterparts in a rather straightforward way.%
\footnote{In general, the higher-derivative vertices are degree-$n$, $n\geq 2$, polynomials in $\Po$. For fields with particular values of spins, some higher-derivative vertices are degree-1 polynomials in $\Po$. For these particular cases, expressions for higher-derivative and first-derivative cubic vertices coincide.
}
Namely, some higher-derivative cubic vertices are obtained by replacement $\kappa=\irm \Po$ in expression for $p_\smpth^-$ \rf{01032020-man-x01} (for some details, see Appendix A). Doing so, we get for some bose and fermi-bose cubic vertices the following higher-derivative representation:
\beq
\label{01032020-man-12}
&& \hspace{-1cm} p_\smpth^- = V_\HDsm \,, \hspace{1cm}  V_\HDsm = V\big|_{\kappa = \irm \Po}\,.
\\
\label{01032020-man-13} && V_\HDsm = V_\Bsm(L_a) \,, \hspace{2cm} \hbox{ for bose vertices},
\\
\label{01032020-man-14} && V_\HDsm = K V_\FBsm(L_a)\,, \hspace{1.4cm} \hbox{ for fermi-bose vertices},
\\
\label{01032020-man-15} && \hspace{1cm} L_a \equiv  \irm \frac{\Po}{\beta_a} + \frac{\betach_a}{2\beta_a} m_a + \frac{m_{a+1}^2 -m_{a+2}^2}{2m_a} \,,
\hspace{0.5cm}  K\equiv \frac{1}{\beta_1\beta_2}(\irm \Po + m_1\beta_2 - m_2\beta_1 \big)\,,\qquad
\eeq
where for fermi-bose vertices, the external line indices $a=1,2$ stand for two fermionic fields entering the cubic vertex. Note that $L_a$ appear in \rf{01032020-man-13},\rf{01032020-man-14} only iff $m_a\ne0$, Expression for $j_\smpth^{-1}$ corresponding to cubic vertex \rf{01032020-man-12} is given by
\beq
\label{01032020-man-17} && j_\smpth^{-1} =  -  \sum_{a=1,2,3} \frac{2\irm \betach_a}{3\beta_a} \partial_{L_a}   p_\smpth^- \,, \hspace{3.3cm} \hbox{ for bose vertices},
\\
\label{01032020-man-18} && j_\smpth^{-1} = \Big(  \frac{\irm \betach_3}{3\beta_1\beta_2}  - K \sum_{a=1,2,3} \frac{2\irm \betach_a}{3\beta_a} \partial_{L_a} \Big) V_\FBsm\,, \hspace{1cm} \hbox{ for fermi-bose vertices}.
\eeq
We emphasize that expressions for $p_\smpth^-$ \rf{01032020-man-13},\rf{01032020-man-14} and for $ j_\smpth^{-1}$ in \rf{01032020-man-17},\rf{01032020-man-18} are valid for some (not all) cubic vertices we discuss below. For the remaining cubic vertices, the higher-derivative form of
$p_\smpth^-$ depends not only on $L_1$, $L_2$, $L_3$ but also on $\beta_1$, $\beta_2$, $\beta_3$. Besides this, we find also a wide class of higher-derivative vertices that are expressed entirely in terms of $L_{\crit,a}$. For some such cubic vertices, the procedure for generating higher-derivative vertices $p_\smpth^-$ is clarified in Appendix A, while the expressions $p_\smpth^-$ and $j_\smpth^{-1}$ are presented explicitly in Sec.\ref{bos-sec-06} and Sec.\ref{fer-bos-sec-07}.
We now apply our result above-presented for discussion of all cubic vertices in turn.

%%%%%%%%%%%%%%%%%%%%%%%%%%%%%%%%%%%%%%%%%%%%%%%%%%%%%%%%%%%%%%%%%%%%%%%%%%%%%%%
%%%%%%%%%%%%%%%%%%%%%%%%%%%%%%%%%%%%%%%%%%%%%%%%%%%%%%%%%%%%%%%%%%%%%%%%%%%%%%%
\newsection{ \large Cubic interaction vertices for bosonic fields} \label{bos-sec-06}
%%%%%%%%%%%%%%%%%%%%%%%%%%%%%%%%%%%%%%%%%%%%%%%%%%%%%%%%%%%%%%%%%%%%%%%%%%%%%%%
%%%%%%%%%%%%%%%%%%%%%%%%%%%%%%%%%%%%%%%%%%%%%%%%%%%%%%%%%%%%%%%%%%%%%%%%%%%%%%%

{\bf Classification of bose vertices}. Cubic vertices describing interaction of three bosonic fields we refer to as bose vertices. Consider bose vertex for three fields having masses $m_1$, $m_2$, $m_3$. To develop classification of bose vertices we introduce a quantities $D$, $\Pbf_{\epsilon m}$ defined by the relations
\beq
\label{02032020-man-01} && D \equiv m_1^4 + m_2^4 + m_3^4 - 2m_1^2m_2^2 -2 m_2^2 m_3^2 - 2 m_3^2 m_1^2\,,
\\
\label{02032020-man-02} && \Pbf_{\epsilon m}\equiv \sum_{a=1,2,3} \epsilon_a m_a  \,, \qquad \epsilon_1^2 =1\,, \quad \epsilon_2^2 =1\,, \quad \epsilon_3^2 =1\,,
\\
\label{02032020-man-03} && \hspace{1cm}D = (m_1 + m_2 + m_3)(m_1 + m_2 - m_3)(m_1 - m_2 + m_3)(m_1 - m_2 - m_3)\,, \qquad
\eeq
where, in \rf{02032020-man-03}, we present helpful alternative representation for the quantity $D$ defined in \rf{02032020-man-01}. From \rf{02032020-man-02}, \rf{02032020-man-03} it is seen that $D\ne 0$ iff $\Pbf_{\epsilon m}\ne 0$ for all admitted values of $\epsilon_1$, $\epsilon_2$, $\epsilon_3$.
Also, from \rf{02032020-man-02}, \rf{02032020-man-03}, we see that, if $\Pbf_{\epsilon m} = 0$ for some values of $\epsilon_1$, $\epsilon_2$, $\epsilon_3$, then $D=0$.

Depending on the masses $m_1$, $m_2$, $m_3$, we split cubic vertices in the following four groups:
\beq
\label{02032020-man-04} && \hspace{-1cm} \hbox{\bf \small Ia})  \hspace{0.5cm}  m_1 \ne 0 \,, \quad m_2 \ne 0 \,, \quad m_3 \ne 0 \,, \hspace{0.7cm} D \ne 0;
\\
\label{02032020-man-05} && \hspace{-1cm} \hbox{\bf \small Ib})  \hspace{0.5cm} m_1\ne0\,,\quad m_2 \ne0\,,\quad m_3\ne 0\,, \hspace{0.7cm} \Pbf_{\epsilon m} = 0\,, \qquad D=0\,;
\\
\label{02032020-man-06} && \hspace{-1cm} \hbox{\bf \small IIa}) \hspace{0.5cm} m_1 \ne 0 \,, \quad m_2\ne 0\,, \quad |m_1|\ne |m_2|\,, \quad m_3=0\,;\qquad
\\
\label{02032020-man-07} && \hspace{-1cm} \hbox{\bf \small IIb})  \hspace{0.5cm} m_1 \ne 0 \,, \quad m_2\ne 0\,, \quad |m_1| = |m_2|\,, \quad m_3 = 0\,;
\\
\label{02032020-man-08} && \hspace{-1cm} \hbox{\bf \small III})  \hspace{0.5cm}  m_1 = 0 \,, \quad m_2= 0\,, \quad  m_3 \ne 0\,;
\\
\label{02032020-man-09} && \hspace{-1cm} \hbox{\bf \small IV})  \hspace{0.5cm}  m_1 = 0 \,, \quad m_2 = 0 \,, \quad m_3 = 0 \,.
\eeq

Now, using our classification in \rf{02032020-man-04}-\rf{02032020-man-09}, we discuss the respective cubic vertices in turn.

\noindent {\bf Ia)  Cubic  vertex for three arbitrary spin massive fields with masses $D\ne 0$}.
Using notation as in \rf{27022020-man-08-a1}-\rf{27022020-man-08-a3}, we consider a cubic vertex for three
fields with the following masses and spins:
\beq
\label{02032020-man-10} && (m_1,s_1)-(m_2,s_2)-(m_3,s_3)\,, \hspace{1cm} s_1,s_2,s_3 \in \No_0\,,
\nonumber\\
\label{02032020-man-11} &&   m_1 \ne 0\,,\quad m_2 \ne 0\,,\quad m_3 \ne 0; \hspace{1cm} D\ne 0\,,
\eeq
where $D$ is defined in \rf{02032020-man-01}. Solution for vertices $V$, $\Vb$ entering cubic vertex $p_\smpth^-$ \rf{01032020-man-x01} is given by
\beq
\label{02032020-man-12} && \hspace{-1cm} V = C V_\kappa\,, \qquad \Vb = C V_{-\kappa}\,,
\\
\label{02032020-man-13} && V_\kappa =   L_{\kappa,1}^{s_1} L_{\kappa,2}^{s_2} L_{\kappa,3}^{s_3}\,, \hspace{1cm} L_{\kappa,a} \equiv  \frac{\kappa}{\beta_a} + \frac{\betach_a}{2\beta_a} m_a + \frac{m_{a+1}^2 -m_{a+2}^2}{2m_a}  \,,
\eeq
where $C$ is coupling constant and $\kappa$ is defined in \rf{01032020-man-x02}.
Plugging $V$ and $\Vb$ \rf{02032020-man-12} into \rf{01032020-man-x01}, we get first-derivative cubic vertex $p_\smpth^-$.
In this paper, unless otherwise specified, {\it all coupling constants are real-valued}. Note also that, in general, our coupling constants might depend on masses and spins of fields entering cubic vertex. Thus, {\it there is only one type \hbox{\bf \small Ia} first-derivative cubic vertex}.

\noindent {\bf Higher-derivative form}. Higher-derivative form of the vertex in \rf{02032020-man-11} is obtained by plugging $\kappa=\irm \Po$ into expressions for $V$, $\Vb$ \rf{02032020-man-12} and $p_\smpth^-$ \rf{01032020-man-x01}. Doing so, we get {\it one higher-derivative cubic vertex}
\be \label{02032020-man-15}
p_\smpth^- =   C L_1^{s_1} L_2^{s_2} L_3^{s_3}\,,
\hspace{1cm} L_a \equiv  \irm \frac{\Po}{\beta_a} + \frac{\betach_a}{2\beta_a} m_a + \frac{m_{a+1}^2 -m_{a+2}^2}{2m_a} \,,
\ee
where the coupling constant $C$ \rf{02032020-man-15} coincides with $C$ in \rf{02032020-man-12}.
Expression for $j_\smpth^{-1}$ can be obtained by plugging $p_\smpth^-$ \rf{02032020-man-15} into \rf{01032020-man-17}.

\noindent {\bf Ib)  Cubic  vertices for three arbitrary spin massive fields with masses $\Pbf_{\epsilon m}=0$}. Using notation in \rf{27022020-man-08-a1}-\rf{27022020-man-08-a3}, we consider a cubic vertices for three
fields with the following masses and spins:
\beq
\label{02032020-man-17} && (m_1,s_1)-(m_2,s_2)-(m_3,s_3)\,, \hspace{1cm} s_1,s_2,s_3 \in \No_0\,,
\nonumber\\
&&  m_1\ne 0\,,\quad m_2\ne 0\,,\quad m_3\ne 0\,, \hspace{1cm}  \Pbf_{\epsilon m} = 0;\qquad D = 0 \,,
\eeq
where $\Pbf_{\epsilon m}$ is defined in \rf{02032020-man-02}. Note that given masses which satisfy the condition $D=0$, the three $\epsilon_a$ \rf{02032020-man-02} are chosen so that to respect the condition $\Pbf_{\epsilon m}=0$.
Given masses and spins, {\it first-derivative form of two cubic vertices} describing interactions of fields in \rf{02032020-man-17} is given by
\beq
\label{02032020-man-19} && \hspace{-1cm} p_\smpth^- =  C_{\epsilon_1\epsilon_2\epsilon_3} ( \irm \Po + \Po_{\epsilon m} )\Po_{\epsilon m}^{\, \Sbf_\epsilon -1} \prod_{a=1,2,3} \beta_a^{-s_{\epsilon a}}\,, \hspace{1cm} \Sbf_\epsilon\in \Zo\,,
\\
\label{02032020-man-19-a} && \hspace{-1cm} p_\smpth^- =  \Cb_{\epsilon_1\epsilon_2\epsilon_3} ( \irm \Po - \Po_{\epsilon m} )\Po_{\epsilon m}^{\, -\Sbf_\epsilon -1} \prod_{a=1,2,3} \beta_a^{s_{\epsilon a}}\,, \hspace{1cm} \Sbf_\epsilon\in \Zo\,,
\\
\label{02032020-man-20} && \hspace{-0.2cm} \Sbf_\epsilon \equiv \sum_{a=1,2,3} s_{\epsilon a}\,, \hspace{2.2cm}  \So_\epsilon \equiv \frac{1}{3}\sum_{a=1,2,3} \betach_a s_{\epsilon a}\,, \hspace{1cm} s_{\epsilon a} \equiv \epsilon_a s_a  \,,
\\
\label{02032020-man-21} && \hspace{-0.2cm} \Po_{\epsilon m} \equiv  \frac{1}{3}\sum_{a=1,2,3} \betach_a \epsilon_a m_a\,,\hspace{0.8cm}   \PP_{\epsilon m} \equiv \sum_{a=1,2,3} \frac{\epsilon_a m_a}{\beta_a}\,,\qquad
\epsilon_1^2=1\,,\ \ \epsilon_2^2=1\,,\ \ \epsilon_3^2=1\,,\qquad
\eeq
where coupling constants $C_{\epsilon_1\epsilon_2\epsilon_3}$, $\Cb_{\epsilon_1\epsilon_2\epsilon_3}$ can depend on masses, spins, and the parameters $\epsilon_1$, $\epsilon_2$, $\epsilon_3$. Vertex \rf{02032020-man-19-a} is obtained from \rf{02032020-man-19} by the replacements $\epsilon_a\rightarrow -\epsilon_a$, i.e., vertices \rf{02032020-man-19},\rf{02032020-man-19-a} describe overcomplete basis of vertices. By using relations \rf{01032020-man-01}, \rf{01032020-man-08}, we find that  expressions for $j_\smpth^-$ corresponding to vertices \rf{02032020-man-19}, \rf{02032020-man-19-a} take the form
\beq
&& \hspace{-1cm} j_\smpth^{-1}  =   2\irm C_{\epsilon_1\epsilon_2\epsilon_3} \Big( - \So_\epsilon \Po_{\epsilon m}   + \frac{\beta}{3}  (\Sbf_\epsilon -1)\PP_{\epsilon m} \Big) \Po_{\epsilon m}^{\Sbf_\epsilon-2} \prod_{a=1,2,3}\beta_a^{-s_{\epsilon a}}\,,
\\
&& \hspace{-1cm} j_\smpth^{-1}  =   2\irm \Cb_{\epsilon_1\epsilon_2\epsilon_3} \Big(  \So_\epsilon \Po_{\epsilon m}   - \frac{\beta}{3}  (\Sbf_\epsilon + 1)\PP_{\epsilon m} \Big) \Po_{\epsilon m}^{-\Sbf_\epsilon-2} \prod_{a=1,2,3}\beta_a^{s_{\epsilon a}}\,.
\eeq

\noindent {\bf Higher-derivative form}.
For $\Sbf_\epsilon>0$ and $\Sbf_\epsilon<0$, the first-derivative cubic vertices \rf{02032020-man-19} and \rf{02032020-man-19-a} admit the following higher-derivative form:
\beq
\label{02032020-man-23-b} &&  \hspace{-1cm}  p_\smpth^- =  C_{\epsilon_1\epsilon_2\epsilon_3}^\HDsm \big( \irm \Po +  \Po_{\epsilon m} \big)^{\Sbf_\epsilon} \prod_{a=1,2,3}\beta_a^{-s_{\epsilon a}}\,,  \hspace{1cm}  \ \Sbf_\epsilon > 0\,, \hspace{0.4cm} C_{\epsilon_1\epsilon_2\epsilon_3}^\HDsm \equiv 2^{1-\Sbf_\epsilon } C_{\epsilon_1\epsilon_2\epsilon_3}\,,
\\
\label{02032020-man-24-b} && \hspace{-1cm} p_\smpth^- =  \Cb_{\epsilon_1\epsilon_2\epsilon_3}^\HDsm \big( \irm \Po -  \Po_{\epsilon m} \big) ^{-\Sbf_\epsilon} \prod_{a=1,2,3}\beta_a^{s_{\epsilon a}}\,,  \hspace{1cm}  \ \Sbf_\epsilon < 0\,, \hspace{0.4cm} \Cb_{\epsilon_1\epsilon_2\epsilon_3}^\HDsm \equiv (-2)^{1+\Sbf_\epsilon } \Cb_{\epsilon_1\epsilon_2\epsilon_3}\,,
\eeq
where we use notation \rf{02032020-man-20}, while $C_{\epsilon_1\epsilon_2\epsilon_3}$, $\Cb_{\epsilon_1\epsilon_2\epsilon_3}$ \rf{02032020-man-23-b},\rf{02032020-man-24-b} coincide with coupling constants in \rf{02032020-man-19},\rf{02032020-man-19-a}.
To find $j_\smpth^{-1}$ we use general relation \rf{29022020-man-17}. Doing so and using notation \rf{02032020-man-20},\rf{02032020-man-21}, we get $j_\smpth^{-1}$ corresponding to $p_\smpth^- $ \rf{02032020-man-23-b}, \rf{02032020-man-24-b},
\beq
\label{02032020-man-24-c1} && j_\smpth^{-1}  =  - 2\irm C_{\epsilon_1\epsilon_2\epsilon_3}^\HDsm \So_\epsilon (\irm \Po + \Po_{\epsilon m})^{\Sbf_\epsilon-1} \prod_{a=1,2,3}\beta_a^{-s_{\epsilon a}}\,, \hspace{1cm} \Sbf_\epsilon > 0\,,
\\
\label{02032020-man-24-c2} && j_\smpth^{-1}  =     2\irm \Cb_{\epsilon_1\epsilon_2\epsilon_3}^\HDsm \So_\epsilon  (\irm \Po- \Po_{\epsilon m})^{-\Sbf_\epsilon-1} \prod_{a=1,2,3}\beta_a^{s_{\epsilon a}}\,,  \hspace{1.5cm} \Sbf_\epsilon < 0\,.
\eeq
Vertices \rf{02032020-man-23-b}, \rf{02032020-man-24-b} turn out to be building blocks for vertices obtained via dimensional reduction from cubic vertices of massless fields in $R^{3,1}$. For details of derivation and our proposal for higher-spin theory of massive fields in $R^{2,1}$, see Appendix B. Note that restrictions on $\Sbf_\epsilon$ in \rf{02032020-man-23-b}-\rf{02032020-man-24-c2} can be ignored only for the case when  $s_1=s_2=s_3=0$.

Note that vertices \rf{02032020-man-19},\rf{02032020-man-19-a} are defined for all $\Sbf_\epsilon \in \Zo$, while in \rf{02032020-man-23-b},\rf{02032020-man-24-b} the values of $\Sbf_\epsilon$ are restricted. In other words, first-derivative vertices \rf{02032020-man-19},\rf{02032020-man-19-a} provide us the full list of vertices, while higher-derivative vertices \rf{02032020-man-23-b},\rf{02032020-man-24-b} provide us the particular list of vertices.

As a remark, we note that vertices \rf{02032020-man-23-b},\rf{02032020-man-24-b} can be represented in terms of quantities $L_{\crit,a}$, $\Lb_{\crit,a}$ defined by
\be \label{02032020-man-24-c3}
L_{\crit,a}  \equiv     \frac{1}{\beta_a}\big( \irm \Po  + \Po_{\epsilon m}\big)\,, \hspace{1cm} \Lb_{\crit,a}  \equiv     \frac{1}{\beta_a}\big( \irm \Po  - \Po_{\epsilon m}\big)\,.
\ee
For example, vertex \rf{02032020-man-23-b} can be represented as $C_{\epsilon_1\epsilon_2\epsilon_3}^\HDsm L_{\crit,1}^{\epsilon_1 s_1} L_{\crit,2}^{\epsilon_2 s_2} L_{\crit,3}^{\epsilon_3 s_3}$.

\noindent {\bf {\small  IIa}) Cubic  vertex for two arbitrary spin massive fields with masses $|m_1|\ne |m_2|$ and one scalar massless field}. Using notation as in \rf{27022020-man-08-a1}-\rf{27022020-man-08-a3}, we consider a cubic vertex for three fields with the following masses and spins:
\beq
\label{02032020-man-25} && (m_1,s_1)-(m_2,s_2)-(0,0)\,, \hspace{1cm} s_1,s_2 \in \No_0\,,
\nonumber\\
\label{02032020-man-26} && m_1 \ne 0,\quad m_2\ne 0,\quad |m_1|\ne |m_2|,
\eeq
Solution for vertices $V$, $\Vb$ entering cubic vertex $p_\smpth^-$ \rf{01032020-man-x01} is given by
\beq
\label{02032020-man-27} && \hspace{-1cm} V = C V_\kappa\,, \qquad \Vb = C V_{-\kappa}\,, %
\\
\label{02032020-man-28} && V_\kappa \equiv   L_{\kappa,1}^{s_1} L_{\kappa,2}^{s_2} \,, \hspace{1cm}  \kappa^2 = - \beta_2 \beta_3 m_1^2 - \beta_1\beta_3 m_2^2\,,
\\
&& L_{\kappa,1} \equiv  \frac{ \kappa}{\beta_1} + \frac{\betach_1}{2\beta_1} m_1 + \frac{m_2^2}{2m_1} \,, \hspace{1cm} L_{\kappa,2} \equiv   \frac{ \kappa}{\beta_2} +\frac{\betach_2}{2\beta_2} m_2 - \frac{m_1^2}{2m_2} \,,
\eeq
where $C$ is a coupling constant. Plugging $V$ and $\Vb$ \rf{02032020-man-27} into \rf{01032020-man-x01}, we get first-derivative form of the cubic vertex $p_\smpth^-$.
Thus, for masses and spins in \rf{02032020-man-26}, there is one vertex.

\noindent {\bf Higher-derivative form}. Higher-derivative form of vertex \rf{02032020-man-26} is obtained by plugging $\kappa=\irm \Po$ into expressions for $V$, $\Vb$ \rf{02032020-man-27} and $p_\smpth^-$ \rf{01032020-man-x01}. Doing so, we get
\be
\label{02032020-man-31}   p_\smpth^- =   C L_1^{s_1} L_2^{s_2}\,,\hspace{0.6cm}
L_1 \equiv    \irm \frac{ \Po}{\beta_1} + \frac{\betach_1}{2\beta_1} m_1 + \frac{m_2^2}{2m_1}  \,, \hspace{0.6cm} L_2 \equiv \irm  \frac{ \Po}{\beta_2} + \frac{\betach_2}{2\beta_2} m_2 - \frac{m_1^2}{2m_2}  \,,
\ee
where the coupling constant $C$ in \rf{02032020-man-31} is the same as the one in \rf{02032020-man-27}. Expression for $j_\smpth^{-1}$ can be obtained by plugging $p_\smpth^-$ \rf{02032020-man-31} into \rf{01032020-man-17}.

\noindent {\bf IIb)  Cubic vertices for two arbitrary spin massive fields with masses $|m_1|=|m_2|$ and one scalar massless field}. Using notation as in \rf{27022020-man-08-a1}-\rf{27022020-man-08-a3}, we consider a cubic vertex for three fields with the following masses and spins:
\beq
\label{02032020-man-34} && (m_1,s_1)-(m_2,s_2)-(m_3,0)\,, \hspace{1.2cm} s_1,s_2 \in \No_0\,,
\nonumber\\
\label{02032020-man-35} &&  m_1\ne 0\,, \quad m_2\ne 0\,, \quad |m_1|=|m_2|\,,\quad m_3 =  0\,.
\eeq
The constraint $|m_1|=|m_2|$ leads to the consideration of two different cases: $m_1=m_2$ and $m_1=-m_2$. For these two cases, we find four vertices which we label as {\small\bf IIb1-IIb4}. First-derivative form of these four vertices is given by
\beq
\label{02032020-man-36} && \hbox{\small\bf IIb1)} \hspace{2cm}  p_\smpth^- =  C_+ (1 + \frac{\irm \Po}{m_1 \beta_3 }) (\frac{\beta_3}{\beta_1})^{-s_1}  (\frac{\beta_3}{\beta_2})^{s_2}\,, \hspace{0.8cm} \hbox{ for } \ m_1=m_2; \hspace{3cm}
\\
\label{02032020-man-37} && \hbox{\small\bf IIb2)} \hspace{2cm} p_\smpth^-  = \Cb_+  (1 - \frac{\irm \Po}{m_1 \beta_3} ) (\frac{\beta_3}{\beta_1})^{s_1}  (\frac{\beta_3}{\beta_2})^{-s_2} \hspace{1cm} \hbox{ for } \ m_1=m_2;
\\
\label{02032020-man-38} && \hbox{\small\bf IIb3)} \hspace{2cm}  p_\smpth^- =  C_- (1 + \frac{\irm \Po}{m_1 \beta_3 }) (\frac{\beta_3}{\beta_1})^{-s_1}  (\frac{\beta_3}{\beta_2})^{-s_2} \hspace{0.8cm} \hbox{ for } \ m_1 = - m_2;
\\
\label{02032020-man-39} && \hbox{\small\bf IIb4)} \hspace{2cm}  p_\smpth^-  = \Cb_-  (1 - \frac{\irm \Po}{m_1 \beta_3} ) (\frac{\beta_3}{\beta_1})^{s_1}  (\frac{\beta_3}{\beta_2})^{s_2} \hspace{1.3cm} \hbox{ for } \ m_1= - m_2\,.
\eeq
We see that for masses $m_1=m_2$ there are two cubic vertices given in \rf{02032020-man-36}, \rf{02032020-man-37}, while,
for masses $m_1=-m_2$, two cubic vertices are given in \rf{02032020-man-38}, \rf{02032020-man-39}. Expressions for $j_\smpth^{-1}$ corresponding to cubic vertices \rf{02032020-man-36}-\rf{02032020-man-39} are obtained by using general relations \rf{01032020-man-01},\rf{01032020-man-08}.

As a remark, we note that all vertices \rf{02032020-man-36}-\rf{02032020-man-39} can be presented on an equal footing as
\beq
\label{02032020-man-39-a1} && \hspace{-0.5cm} p_\smpth^- = C_{\epsilon_1\epsilon_2} \big(1 + \frac{\irm \epsilon_2 \Po}{m_2 \beta_3} \big) \big(\frac{\beta_3}{\beta_1}\big)^{\epsilon_1 s_1}
\big(\frac{\beta_3}{\beta_2}\big)^{\epsilon_2 s_2}\,, \hspace{0.4cm} \epsilon_1 m_1 + \epsilon_2 m_2 =0 \,, \hspace{0.4cm} \epsilon_1^2=1\,, \quad \epsilon_2^2=1\,,\qquad
\\
\label{02032020-man-39-a2}&& \hspace{-0.5cm} C_{-11} = C_+\,, \qquad C_{1-1} = \Cb_+\,, \qquad C_{-1-1} = C_-\,, \qquad
C_{11} = \Cb_-\,,
\eeq
where, in \rf{02032020-man-39-a2}, we identify coupling constants in \rf{02032020-man-36}-\rf{02032020-man-39} and \rf{02032020-man-39-a1}.

\noindent {\bf Higher-derivative form}. For the cases {\small\bf IIb1,IIb2,IIb4}, we find the following higher-derivative representation,
\beq
\label{02032020-man-40} && \hspace{-2cm} \hbox{\small\bf IIb1)} \hspace{0.6cm} p_\smpth^- =  C_+^\HDsm   (\frac{\beta_3}{\beta_1})^{-s_1}  L_2^{s_2}\,, \hspace{0.7cm}  C_+^\HDsm = \frac{2 C_+}{(2m_2)^{s_2}}\,, \hspace{2.3cm} m_1=m_2\,,
\\
\label{02032020-man-41} && \hspace{-2cm} \hbox{\small\bf IIb2)} \hspace{0.6cm} p_\smpth^-  = \Cb_+^\HDsm  L_1^{s_1}  (\frac{\beta_3}{\beta_2})^{-s_2}\,, \hspace{0.7cm}  \Cb_+^\HDsm = \frac{2 \Cb_+}{(-2m_2)^{s_1}}\,, \hspace{2cm} m_1=m_2\,,
\\
\label{02032020-man-42} && \hspace{-2cm} \hbox{\small\bf IIb4)} \hspace{0.6cm} p_\smpth^- =  \Cb_-^\HDsm   L_1^{s_1}  L_2^{s_2}\,, \hspace{1.5cm}  \Cb_-^\HDsm = \frac{2\Cb_-}{(2m_2)^{s_1+s_2}} \,, \hspace{1.7cm} m_1=-m_2\,,
\\
&& \hspace{2cm} L_1 \equiv     \frac{1}{\beta_1}\big(\irm \Po - m_1 \beta_3\big)\,,
\hspace{1cm} L_2 \equiv     \frac{1}{\beta_2}\big(\irm \Po + m_2 \beta_3\big)\,,
\eeq
where, in  \rf{02032020-man-40}-\rf{02032020-man-42}, we also show interrelations  between coupling constants of higher-derivative vertices and coupling constants of first-derivative vertices \rf{02032020-man-36}-\rf{02032020-man-39}.
Expression for $j_\smpth^{-1}$ corresponding to $p_\smpth^-$ \rf{02032020-man-42} can be obtained by plugging $p_\smpth^-$ \rf{02032020-man-42} into \rf{01032020-man-17}, while expressions for $j_\smpth^{-1}$ corresponding to $p_\smpth^-$ in \rf{02032020-man-40},\rf{02032020-man-41} are given by
\beq
\label{02032020-man-42-a1}&&  \hspace{-2cm} \hbox{\small\bf IIb1)} \hspace{0.6cm}  j_\smpth^{-1} =  C_+^\HDsm \Big(  2\irm s_1 - \frac{2\irm \betach_2}{3\beta_2}s_2 \Big) (\frac{\beta_3}{\beta_1})^{-s_1} L_2^{s_2-1}\,, \hspace{1.2cm} m_1=m_2\,, \hspace{0.5cm} s_2 \geq 1;
\\
\label{02032020-man-42-a2} &&  \hspace{-2cm} \hbox{\small\bf IIb2)} \hspace{0.6cm}  j_\smpth^{-1} =  - \Cb_+^\HDsm  \Big(  2\irm s_2 + \frac{2\irm \betach_1}{3\beta_1}s_1 \Big) L_1^{s_1-1} (\frac{\beta_3}{\beta_2})^{-s_2}\,, \hspace{0.8cm} m_1=m_2\,, \hspace{0.6cm} s_1 \geq 1\,.
\eeq
Restrictions $s_2\geq 1$ \rf{02032020-man-42-a1} and $s_1\geq 1$ \rf{02032020-man-42-a2} are obtained by requiring that the $j_\smpth^{-1}$ given in \rf{02032020-man-42-a1}, \rf{02032020-man-42-a2} be polynomial in $L_2$ and $L_1$ respectively. These restrictions are redundant for $j_\smpth^{-1}$ corresponding to all first-derivative vertices in \rf{02032020-man-36}-\rf{02032020-man-39}. Thus, we see that first-derivative vertices \rf{02032020-man-36}-\rf{02032020-man-39} provide us the full list of vertices, while higher-derivative vertices \rf{02032020-man-40}-\rf{02032020-man-42} provide us the particular list of vertices.

As a remark, we note that, by using $L_{\crit, a}$ \rf{02032020-man-24-c3}, we find alternative higher-derivative form for some first-derivative vertices \rf{02032020-man-36}-\rf{02032020-man-39} given by
\be \label{02032020-man-42-a3}
p_\smpth^- =  C_{\epsilon_1\epsilon_2}^\HDsm (\irm \Po +\Po_{\epsilon m})^{\Sbf_\epsilon} \beta_1^{-\epsilon_1 s_1 } \beta_2^{-\epsilon_2 s_2 } \,, \hspace{1cm} \Po_{\epsilon m} = \epsilon_2 m_2 \beta_3 \hspace{1cm}  C_{\epsilon_1\epsilon_2}^\HDsm =  \frac{2 C_{\epsilon_1\epsilon_2}}{(2\epsilon_2 m_2)^{\Sbf_\epsilon}}\,,\qquad
\ee
where $\epsilon_1^2=\epsilon_2^2=1$, $\Sbf_\epsilon= \epsilon_1 s_1 + \epsilon_2 s_2$.
Vertices \rf{02032020-man-42-a3} and corresponding $j_\smpth^{-1}$ are obtained by setting $m_3=0$, $s_3=0$ in \rf{02032020-man-23-b} and \rf{02032020-man-24-c1}. This implies the restriction $\Sbf_\epsilon > 0$ in \rf{02032020-man-42-a3} which is redundant for first-derivative vertices \rf{02032020-man-39-a1}.
Thus, higher-derivative vertices \rf{02032020-man-42-a3} provide us the particular vertices, while first-derivative vertices \rf{02032020-man-36}-\rf{02032020-man-39} provide us the full list of vertices.

\noindent {\bf{\small  III}) Cubic  vertex for two scalar massless fields and one arbitrary spin massive field}. Using notation as in \rf{27022020-man-08-a1}-\rf{27022020-man-08-a3}, we consider a cubic vertex for three fields with the following masses and spins:
\be \label{02032020-man-43}
(0,0)-(0,0)-(m_3,s_3)\,, \hspace{1cm}  m_3 \ne 0\,, \hspace{1cm} s_3 \in \No_0\,.
\ee
Solution for vertices $V$, $\Vb$ entering cubic vertex $p_\smpth^-$ \rf{01032020-man-x01} is given by
\beq
\label{02032020-man-45} && \hspace{-1cm} V = C V_\kappa\,, \qquad \Vb = C V_{-\kappa}\,,
\\
\label{02032020-man-46} && V_\kappa \equiv L_{\kappa,3}^{s_3}\,, \hspace{1cm}  L_{\kappa,3} \equiv   \frac{ \kappa}{\beta_3} + \frac{\betach_3}{2\beta_3} m_3  \,, \hspace{1cm} \kappa^2 = -\beta_1\beta_2 m_3^2\,,
\eeq
where $C$ is a coupling constant. Plugging $V$, $\Vb$ \rf{02032020-man-45} into \rf{01032020-man-x01}, we get first-derivative form of the cubic vertex $p_\smpth^-$.

\noindent {\bf Higher-derivative form}. Higher-derivative form of the cubic vertex in \rf{02032020-man-43} is obtained by plugging $\kappa=\irm \Po$ into expressions for $V$, $\Vb$ \rf{02032020-man-45} and $p_\smpth^-$ \rf{01032020-man-x01}. Doing so, we get
\be \label{02032020-man-48}
p_\smpth^- =   C L_3^{s_3}\,, \hspace{1cm} L_3 \equiv  \irm \frac{ \Po}{\beta_3} + \frac{\betach_3}{2\beta_3} m_3  \,,
\ee
where the coupling constant $C$ in \rf{02032020-man-48} is the same as the one in \rf{02032020-man-45}.  Expression for $j_\smpth^{-1}$ corresponding to $p_\smpth^-$ \rf{02032020-man-48} can be obtained by plugging $p_\smpth^-$ \rf{02032020-man-48} into \rf{01032020-man-17}.

\noindent {\bf IV) Cubic  vertices for three scalar massless fields}. Using notation as in \rf{27022020-man-08-a1}-\rf{27022020-man-08-a3}, we consider a cubic vertex for three massless scalar fields:
\be \label{02032020-man-50}
(0,0)-(0,0)-(0,0)\,.
\ee
For this case $\kappa=0$.  Transformation \rf{01032020-man-11} is not invertible and we use the vertices $V_0$, $V_1$ which should satisfy equations \rf{01032020-man-09} and equations obtained by setting $m_1=m_2=m_3=0$ in \rf{01032020-man-06},\rf{01032020-man-07}. For $m_1=m_2=m_3=0$, equation \rf{01032020-man-06} is satisfied automatically. Thus, for vertices $V_0$ and $V_1$, we get equations \rf{01032020-man-09} and equation $\No_\beta V_0=0$. Solution to these equations is given by
\be \label{02032020-man-52}
V_0 =  C_0 \,, \hspace{1cm} V_1 = \frac{1}{\beta_3} v\big( \frac{\beta_1}{\beta_2}\big)\,,
\ee
where $C_0$ is coupling constant, while $v$ is $\beta$-analytic in $\beta_1/\beta_2$.

We recall that, for $\kappa\equiv\hspace{-0.5cm}/\hspace{0.3cm} 0$, we dealt with one or two vertices. We now see that, for three massless fields, $\kappa\equiv 0$, we have infinite number of light-cone gauge vertices \rf{02032020-man-52} parametrized by the constant $C_0$ and by the function $v$. The vertex $V_0=C_0$ is associated with the standard $\phi^3$ self-interaction of scalar field in Lorentz covariant approach, while vertex $V_1$ \rf{02032020-man-52} is associated with the Yang-Mills cubic interaction, when $v=const$ . To our knowledge, Lorentz covariant vertices associated with our light-cone gauge vertex $V_1$, when $v \ne const$, are not available in the literature. In other words, we encounter a mismatch between classifications of light-cone gauge and Lorentz covariant cubic vertices.
For massless field in $4d$, the mismatch between classifications of light-cone gauge vertices and Lorentz covariant cubic vertices is well known (see, e.g., Ref.\cite{Conde:2016izb}).

%%%%%%%%%%%%%%%%%%%%%%%%%%%%%%%%%%%%%%%%%%%%%%%%%%%%%%%%%%%%%%%%%%%%%%%%%%%%%%%
%%%%%%%%%%%%%%%%%%%%%%%%%%%%%%%%%%%%%%%%%%%%%%%%%%%%%%%%%%%%%%%%%%%%%%%%%%%%%%%
\newsection{ \large Cubic interaction vertices for fermonic and bosonic fields} \label{fer-bos-sec-07}
%%%%%%%%%%%%%%%%%%%%%%%%%%%%%%%%%%%%%%%%%%%%%%%%%%%%%%%%%%%%%%%%%%%%%%%%%%%%%%%
%%%%%%%%%%%%%%%%%%%%%%%%%%%%%%%%%%%%%%%%%%%%%%%%%%%%%%%%%%%%%%%%%%%%%%%%%%%%%%%

{\bf Classification of fermi-bose vertices}. Cubic vertices describing interaction of two fermionic fields and one bosonic field we refer to as fermi-bose vertices.
Our conventions for fields in the cubic vertex are as follows. In the bose-fermi vertices, two fermionic fields carry external line indices $a=1,2$,
while one bosonic field corresponds to $a=3$. Namely, using notation as in \rf{27022020-man-08-a1}-\rf{27022020-man-08-a6}, in the bose-fermi vertices, two fermionic fields and one bosonic field are identified as
\be
(m_1,s_1+\half), (m_2,s_2+\half) - \hbox{two fermionic fields}\,, \qquad  (m_3,s_3)- \hbox{one bosonic field}.\qquad
\ee
As before, to develop classification of fermi-bose vertices we use the quantities $
D$, $\Pbf_{\epsilon m}$ defined in \rf{02032020-man-01}\rf{02032020-man-02}.
Namely,
depending on the masses of two fermionic fields $m_1$, $m_2$, and mass of one bosonic field $m_3$, we split vertices in the following four groups:
\beq
\label{03032020-man-01} && \hspace{-1cm} \hbox{\bf \small Ia})  \hspace{0.7cm}  m_1 \ne 0 \,, \quad m_2 \ne 0 \,, \quad m_3 \ne 0 \,, \hspace{0.7cm} D \ne 0;
\\
\label{03032020-man-02} && \hspace{-1cm} \hbox{\bf \small Ib})  \hspace{0.7cm} m_1\ne0\,,\quad m_2 \ne0\,,\quad m_3\ne 0\,, \hspace{0.7cm} \Pbf_{\epsilon m} = 0, \qquad D=0;
\\
\label{03032020-man-03} && \hspace{-1cm} \hbox{\bf \small IIa1}) \hspace{0.5cm} m_1 \ne 0 \,, \quad m_2\ne 0\,, \quad |m_1|\ne |m_2|\,, \quad m_3=0\,;\qquad
\\
\label{03032020-man-04} && \hspace{-1cm} \hbox{\bf \small IIa2}) \hspace{0.5cm} m_1 = 0 \,, \quad m_2\ne 0\,, \quad m_3 \ne 0\,, \qquad  \quad |m_2| \ne |m_3|\,,\qquad
\\
\label{03032020-man-05} && \hspace{-1cm} \hbox{\bf \small IIb1})  \hspace{0.5cm} m_1 \ne 0 \,, \quad m_2\ne 0\,, \quad |m_1| = |m_2|\,, \quad m_3 = 0\,;
\\
\label{03032020-man-06} && \hspace{-1cm} \hbox{\bf \small IIb2})  \hspace{0.5cm} m_1 = 0 \,, \quad m_2\ne 0\,, \quad m_3 \ne 0\, \quad |m_2| = |m_3|\,;
\\
\label{03032020-man-07} && \hspace{-1cm} \hbox{\bf \small IIIa})  \hspace{0.5cm}  m_1 = 0 \,, \quad m_2= 0\,, \quad  m_3 \ne 0\,;
\\
\label{03032020-man-08}  && \hspace{-1cm} \hbox{\bf \small IIIb})  \hspace{0.5cm}  m_1 \ne 0 \,, \quad m_2= 0\,, \quad  m_3 = 0\,;
\\
\label{03032020-man-09} && \hspace{-1cm} \hbox{\bf \small IV})  \hspace{0.7cm}  m_1 = 0 \,, \quad m_2 = 0 \,, \quad m_3 = 0 \,.
\eeq

Now, using our classification in \rf{03032020-man-01}-\rf{03032020-man-09}, we discuss the respective cubic vertices in turn.

\noindent {\bf Ia)  Cubic  vertex for two fermionic arbitrary spin massive fields and one bosonic arbitrary spin massive field with masses $D\ne 0$}.
Using notation as in \rf{27022020-man-08-a1}-\rf{27022020-man-08-a6}, we consider a cubic vertex for three fields with the following masses and spins:
\beq
\label{03032020-man-10} && (m_1,s_1+\half)-(m_2,s_2+\half)-(m_3,s_3)\,, \hspace{1cm} s_1,s_2,s_3 \in \No_0\,,
\nonumber\\
\label{03032020-man-11} &&   m_1 \ne 0\,,\quad m_2 \ne 0\,,\quad m_3 \ne 0; \hspace{1cm} D\ne 0\,,
\eeq
where $D$ is defined in \rf{02032020-man-01}. Solution for vertices $V$, $\Vb$ entering cubic vertex $p_\smpth^-$ \rf{01032020-man-x01} is given by
\beq
\label{03032020-man-12} && \hspace{-1cm} V = C V_\kappa\,, \qquad \Vb = C V_{-\kappa}\,,
\\
\label{03032020-man-13} && V_\kappa =  K_\kappa L_{\kappa,1}^{s_1} L_{\kappa,2}^{s_2} L_{\kappa,3}^{s_3}\,,
\hspace{1cm} K_\kappa \equiv \frac{1}{\beta_1\beta_2}(\kappa+ m_1\beta_2 -m_2\beta_1)\,,
\\
\label{03032020-man-14} && L_{\kappa,a} \equiv   \frac{ \kappa}{\beta_a} + \frac{\betach_a}{2\beta_a} m_a + \frac{m_{a+1}^2 -m_{a+2}^2}{2m_a} \,, \qquad \kappa^2  = -\beta \sum_{a=1,2,3} \frac{m_a^2}{\beta_a}\,,
\eeq
where $C$ is coupling constant. Plugging $V$, $\Vb$ \rf{03032020-man-12} into \rf{01032020-man-x01}, we get first-derivative form of the cubic vertex $p_\smpth^-$.
We recall that throughout this paper all coupling constants are real valued and, in  general, our coupling constants might depend on masses and spins of fields entering cubic vertex.

\noindent {\bf Higher-derivative form}. Higher-derivative form of the vertex in \rf{03032020-man-11} is obtained by plugging $\kappa=\irm \Po$ into expressions for $V$, $\Vb$ \rf{03032020-man-12} and $p_\smpth^-$ \rf{01032020-man-x01}. Doing so, we get
\beq
\label{03032020-man-15} && p_\smpth^- =   C K L_1^{s_1} L_2^{s_2} L_3^{s_3}\,, \hspace{1cm} K \equiv \frac{1}{\beta_1\beta_2}(\irm \Po + m_1\beta_2 -m_2\beta_1)\,,
\\
\label{03032020-man-16} && L_a \equiv  \irm \frac{\Po}{\beta_a} + \frac{\betach_a}{2\beta_a} m_a + \frac{m_{a+1}^2 -m_{a+2}^2}{2m_a}  \,,
\eeq
where the coupling constant $C$ in \rf{03032020-man-15} is the same as the one in \rf{03032020-man-12}.  Expression for $j_\smpth^{-1}$ corresponding to $p_\smpth^-$ \rf{03032020-man-15} can be obtained by using $p_\smpth^-$ \rf{03032020-man-15} and relations in \rf{01032020-man-12},\rf{01032020-man-14},\rf{01032020-man-18}.

\noindent {\bf Ib)  Cubic  vertices for two fermionic arbitrary spin massive fields and one bosonic arbitrary spin massive field with masses $\Pbf_{\epsilon m}=0$}. Using notation as in \rf{27022020-man-08-a1}-\rf{27022020-man-08-a6}, we consider a cubic vertex for three fields with the following masses and spins:
\beq
\label{03032020-man-17} && (m_1,s_1+\half)-(m_2,s_2+\half)-(m_3,s_3)\,, \hspace{1cm} s_1,s_2,s_3 \in \No_0\,,
\nonumber\\
\label{03032020-man-18} &&  m_1\ne 0\,,\quad m_2\ne 0\,,\quad m_3\ne 0\,, \hspace{1cm}  \Pbf_{\epsilon m} = 0;\qquad
\eeq
where $\Pbf_{\epsilon m}$ is defined in \rf{02032020-man-02}.
First-derivative form of two cubic vertices describing interactions of fields in \rf{03032020-man-17} is given by
\beq
\label{03032020-man-19} && \hspace{-1cm} p_\smpth^- =  C_{\epsilon_1\epsilon_2\epsilon_3}  ( \irm \Po + \Po_{\epsilon m} )\Po_{\epsilon m}^{\, \Sbf_\epsilon^\f -1} \beta_1^{-\half}\beta_2^{-\half} \prod_{a=1,2,3} \beta_a^{-s_{\epsilon a}^\f}\,, \hspace{1cm} \Sbf_\epsilon^\f \in
\Zo\,,
\\
\label{03032020-man-19-a} && \hspace{-1cm} p_\smpth^- =  \Cb_{\epsilon_1\epsilon_2\epsilon_3}  ( \irm \Po - \Po_{\epsilon m} )\Po_{\epsilon m}^{\,- \Sbf_\epsilon^\f -1} \beta_1^{-\half}\beta_2^{-\half} \prod_{a=1,2,3} \beta_a^{s_{\epsilon a}^\f}\,,  \hspace{1cm} \Sbf_\epsilon^\f \in
\Zo\,,
\\
\label{03032020-man-20} && \Sbf^\f_\epsilon \equiv \sum_{a=1,2,3} s_{\epsilon a}^\f\,, \hspace{1cm} s_{\epsilon a}^\f \equiv \epsilon_a s_a^\f\,,\quad s_1^\f \equiv s_1 + \half\,,\quad s_2^\f \equiv s_2+\half\,,\quad s_3^\f \equiv s_3\,,\qquad
\eeq
where $\Po_{\epsilon m}$  and $\epsilon_a$ are defined in \rf{02032020-man-21}.
Coupling constants $C_{\epsilon_1\epsilon_2\epsilon_3}$, $\Cb_{\epsilon_1\epsilon_2\epsilon_3}$ can depend on masses, spins, and the parameters $\epsilon_a$. Vertex \rf{03032020-man-19-a} is obtained from \rf{03032020-man-19} by the replacements $\epsilon_a\rightarrow -\epsilon_a$, i.e., vertices \rf{03032020-man-19},\rf{03032020-man-19-a} describe overcomplete basis of vertices.

\noindent {\bf Higher-derivative form}.  For $\Sbf_\epsilon^\f > 0$ and $\Sbf_\epsilon^\f < 0$, the cubic vertices \rf{03032020-man-19}  \rf{03032020-man-19-a} admit the following alternative higher-derivative representation:
\beq
\label{03032020-man-23-b1} && \hspace{-1cm} p_\smpth^- =  C_{\epsilon_1\epsilon_2\epsilon_3}^\HDsm \big( \irm \Po +  \Po_{\epsilon m} \big)^{\Sbf_\epsilon^\f} \beta_1^{-\half}\beta_2^{-\half} \prod_{a=1,2,3}\beta_a^{-s_{\epsilon a}^\f}\,,  \hspace{0.6cm}  \Sbf_\epsilon^\f > 0\,, \hspace{0.6cm} C_{\epsilon_1\epsilon_2\epsilon_3}^\HDsm = 2^{1-\Sbf_\epsilon^\f} C_{\epsilon_1\epsilon_2\epsilon_3}\,,\qquad
\\
\label{03032020-man-23-b2} && \hspace{-1cm} p_\smpth^- =  \Cb_{\epsilon_1\epsilon_2\epsilon_3}^\HDsm \big( \irm \Po -  \Po_{\epsilon m} \big)^{-\Sbf_\epsilon^\f} \beta_1^{-\half}\beta_2^{-\half} \prod_{a=1,2,3}\beta_a^{s_{\epsilon a}^\f}\,,  \hspace{0.6cm}  \Sbf_\epsilon^\f < 0\,, \hspace{0.6cm} \Cb_{\epsilon_1\epsilon_2\epsilon_3}^\HDsm = (-2)^{1+\Sbf_\epsilon^\f} \Cb_{\epsilon_1\epsilon_2\epsilon_3}\,,\qquad
\eeq
where, in \rf{03032020-man-23-b1},\rf{03032020-man-23-b2}, $C_{\epsilon_1\epsilon_2\epsilon_3}$, $\Cb_{\epsilon_1\epsilon_2\epsilon_3}$ are the  coupling constants entering first-derivative vertices \rf{03032020-man-19}, \rf{03032020-man-19-a}.
Expressions for $j_\smpth^{-1}$ corresponding to $ p_\smpth^- $ \rf{03032020-man-23-b1} and \rf{03032020-man-23-b2} are given by
\beq
\label{03032020-man-23-b3} && j_\smpth^{-1}   =    - 2\irm C_{\epsilon_1\epsilon_2\epsilon_3}^\HDsm   \So_\epsilon^\f  \big(\irm \Po + \Po_{\epsilon m}\big)^{\Sbf_\epsilon^\f-1} \beta_1^{-\half}\beta_2^{-\half} \prod_{a=1,2,3}\beta_a^{-s_{\epsilon a}^\f}\,,
\\
&& j_\smpth^{-1}  =     2\irm \Cb_{\epsilon_1\epsilon_2\epsilon_3}^\HDsm    \So_\epsilon^\f    \big(\irm \Po - \Po_{\epsilon m}\big)^{-\Sbf_\epsilon^\f-1} \beta_1^{-\half}\beta_2^{-\half} \prod_{a=1,2,3}\beta_a^{s_{\epsilon a}^\f}\,,
\eeq
where $\So_\epsilon^\f \equiv \frac{1}{3} \sum_{a=1,2,3}\betach_a s_{\epsilon a}^\f$ and we use the notation as in \rf{03032020-man-20}.

\noindent {\bf {\small  IIa1}) Cubic  vertex for two fermionic arbitrary spin massive fields with masses $|m_1|\ne |m_2|$ and one scalar massless field}. Using notation as in \rf{27022020-man-08-a1}-\rf{27022020-man-08-a6}, we consider a cubic vertex for three fields with the following masses and spins:
\beq
\label{03032020-man-25} && (m_1,s_1+\half)-(m_2,s_2+\half)-(m_3,0)\,, \hspace{1cm} s_1,s_2 \in \No_0\,,
\nonumber\\
\label{03032020-man-26} && m_1 \ne 0,\quad m_2\ne 0,\quad |m_1|\ne |m_2|,\quad m_3=0\,.
\eeq
Solution for vertices $V$, $\Vb$ entering cubic vertex $p_\smpth^-$ \rf{01032020-man-x01} is given by
\beq
\label{03032020-man-27} && \hspace{-1cm} V = C  V_\kappa\,, \qquad \Vb = C V_{-\kappa}\,,
\\
\label{03032020-man-28} && V_\kappa \equiv  K_\kappa L_{\kappa,1}^{s_1} L_{\kappa,2}^{s_2} \,, \hspace{1cm} K_\kappa \equiv \frac{1}{\beta_1\beta_2}(\kappa+ m_1\beta_2 -m_2\beta_1)
\\
\label{03032020-man-29} && L_{\kappa,1} \equiv  \frac{ \kappa}{\beta_1} + \frac{\betach_1}{2\beta_1} m_1 + \frac{m_2^2}{2m_1} \,, \hspace{1.5cm} L_{\kappa,2} \equiv  \frac{ \kappa}{\beta_2} + \frac{\betach_2}{2\beta_2} m_2 - \frac{m_1^2}{2m_2}  \,,
\\
\label{03032020-man-30} &&  \kappa^2 = - \beta_2 \beta_3 m_1^2 - \beta_1\beta_3 m_2^2\,,
\eeq
where $C$ is a coupling constant. Plugging $V$ and $\Vb$ \rf{03032020-man-27} into \rf{01032020-man-x01}, we get first-derivative form of the cubic vertex $p_\smpth^-$.

\noindent {\bf Higher-derivative form}. Higher-derivative form of the vertex in \rf{03032020-man-26} is obtained by plugging $\kappa=\irm \Po$ into expressions for $V$, $\Vb$ \rf{03032020-man-27} and $p_\smpth^-$ \rf{01032020-man-x01}. Doing so, we get
\beq
\label{03032020-man-31} && \hspace{-1cm} p_\smpth^- =   C K L_1^{s_1} L_2^{s_2}\,, \hspace{1cm}  K \equiv \frac{1}{\beta_1\beta_2}(\irm \Po + m_1\beta_2 -m_2\beta_1)\,,
\\
\label{03032020-man-34} && L_1 \equiv    \irm \frac{ \Po}{\beta_1} + \frac{\betach_1}{2\beta_1} m_1 + \frac{m_2^2}{2m_1}  \,, \hspace{1cm} L_2 \equiv   \irm  \frac{ \Po}{\beta_2} + \frac{\betach_2}{2\beta_2} m_2 - \frac{m_1^2}{2m_2}  \,,
\eeq
where the coupling constant $C$ in \rf{03032020-man-31} is the same as the one in \rf{03032020-man-27}. Expression for $j_\smpth^{-1}$ corresponding to $p_\smpth^-$ \rf{03032020-man-31} are obtainable by using $p_\smpth^-$ \rf{03032020-man-31} and general relations in \rf{01032020-man-12},\rf{01032020-man-14},\rf{01032020-man-18}.

\noindent {\bf {\small  IIa2}) Cubic  vertex for one fermionic massless field, one fermionic arbitrary spin massive field and one bosonic arbitrary spin massive field with masses $|m_2|\ne |m_3|$}. Using notation as in \rf{27022020-man-08-a1}-\rf{27022020-man-08-a6}, we consider a cubic vertex for three fields with the following masses and spins:
\beq
\label{03032020-man-34} && (m_1,\half)-(m_2,s_2+\half)-(m_3,s_3+\half)\,, \hspace{1cm} s_2,s_3 \in \No_0\,,
\nonumber\\
\label{03032020-man-35} && m_1 =0,\quad m_2\ne 0,\quad \quad m_3 \ne 0 \,, \quad |m_2|\ne |m_3|\,.
\eeq
Solution for vertices $V$, $\Vb$ entering cubic vertex $p_\smpth^-$ \rf{01032020-man-x01} is given by
\beq
\label{03032020-man-36} && \hspace{-1cm} V = C  V_\kappa\,, \qquad \Vb = C V_{-\kappa}\,,
\\
\label{03032020-man-37} && V_\kappa \equiv  K_\kappa L_{\kappa,2}^{s_2} L_{\kappa,3}^{s_3} \,, \hspace{1cm}  K_\kappa \equiv \frac{1}{\beta_1\beta_2}(\kappa - m_2\beta_1)\,,
\\
&& L_{\kappa,2} \equiv   \frac{ \kappa}{\beta_2} + \frac{\betach_2}{2\beta_2} m_2 + \frac{m_3^2}{2m_2}  \,, \hspace{1cm} L_{\kappa,3} \equiv  \frac{ \kappa}{\beta_3} + \frac{\betach_3}{2\beta_3} m_3 - \frac{m_2^2}{2m_3}  \,,
\\
&&  \kappa^2 = - \beta_1 \beta_3 m_2^2 - \beta_1\beta_2 m_3^2\,,
\eeq
where $C$ is a coupling constant. Plugging $V$, $\Vb$ \rf{03032020-man-36} into \rf{01032020-man-x01}, we get first-derivative form of the cubic vertex $p_\smpth^-$.

\noindent {\bf Higher-derivative form}. Higher-derivative form of the vertex in \rf{03032020-man-35} is obtained by plugging $\kappa=\irm \Po$ into expressions for $V$, $\Vb$ \rf{03032020-man-36} and $p_\smpth^-$ \rf{01032020-man-x01}. Doing so, we get
\beq
\label{03032020-man-38} && \hspace{-1cm} p_\smpth^- =   C K L_2^{s_2} L_3^{s_3}\,, \hspace{1cm}   K \equiv \frac{1}{\beta_1\beta_2}(\irm \Po   - m_2\beta_1)\,,
\\
\label{03032020-man-39} && L_2 \equiv   \irm \frac{ \Po}{\beta_2} + \frac{\betach_2}{2\beta_2} m_2 + \frac{m_3^2}{2m_2}  \,, \hspace{1cm} L_3 \equiv  \irm \frac{ \Po}{\beta_3} +  \frac{\betach_3}{2\beta_3} m_3 - \frac{m_2^2}{2m_3}  \,,
\eeq
where the coupling constant $C$ in \rf{03032020-man-38} is the same as the one in \rf{03032020-man-36}. Expression for $j_\smpth^{-1}$ corresponding to $p_\smpth^-$ \rf{03032020-man-38} can be obtained by using $p_\smpth^-$ \rf{03032020-man-38} and general relations in \rf{01032020-man-12},\rf{01032020-man-14},\rf{01032020-man-18}.

\noindent {\bf IIb1)  Cubic vertices for two fermionic arbitrary spin massive fields with masses $|m_1|=|m_2|$ and one scalar massless field}. Using notation as in \rf{27022020-man-08-a1}-\rf{27022020-man-08-a6}, we consider a cubic vertex for three fields with the following masses and spins:
\beq
&& (m_1,s_1+\half)-(m_2,s_2+\half)-(m_3,0)\,, \hspace{1.2cm} s_1,s_2 \in \No_0\,,
\nonumber\\
\label{03032020-man-42} &&  m_1\ne 0\,, \quad m_2\ne 0\,, \quad |m_1|=|m_2|\,,\quad m_3 =  0\,.
\eeq
The constraint $|m_1|=|m_2|$ leads to the consideration of two different cases: $m_1=m_2$ and $m_1=-m_2$. For these two cases, we find four vertices which we label to as {\small\bf IIb1.1-IIb1.4}. First-derivative form of these four vertices is given by
\beq
\label{03032020-man-43} && \hbox{\small\bf IIb1.1)} \hspace{1.5cm}  p_\smpth^- =  C_+ \frac{1}{\beta_2} (1 + \frac{\irm \Po}{m_1 \beta_3 }) (\frac{\beta_3}{\beta_1})^{-s_1}  (\frac{\beta_3}{\beta_2})^{s_2}\,, \hspace{1.5cm} m_1=m_2; \hspace{3cm}
\\
\label{03032020-man-44} && \hbox{\small\bf IIb1.2)} \hspace{1.5cm} p_\smpth^-  = \Cb_+  \frac{1}{\beta_1} (1 - \frac{\irm \Po}{m_1 \beta_3} ) (\frac{\beta_3}{\beta_1})^{s_1}  (\frac{\beta_3}{\beta_2})^{-s_2}\,, \hspace{1.5cm}   m_1=m_2;
\\
\label{03032020-man-45} && \hbox{\small\bf IIb1.3)} \hspace{1.5cm}  p_\smpth^- =  C_- \frac{1}{\beta_3} (1 + \frac{\irm \Po}{m_1 \beta_3 }) (\frac{\beta_3}{\beta_1})^{-s_1}  (\frac{\beta_3}{\beta_2})^{-s_2}\,, \hspace{1.2cm}  m_1 = - m_2;
\\
\label{03032020-man-46} && \hbox{\small\bf IIb1.4)} \hspace{1.5cm}  p_\smpth^-  = \Cb_-  \frac{\beta_3}{\beta_1\beta_2} (1 - \frac{\irm \Po}{m_1 \beta_3} ) (\frac{\beta_3}{\beta_1})^{s_1}  (\frac{\beta_3}{\beta_2})^{s_2}\,, \hspace{1.3cm} m_1= - m_2\,.
\eeq
Thus, for masses $m_1=m_2$, we find two cubic vertices given in \rf{03032020-man-43}, \rf{03032020-man-44}, while,
for masses $m_1=-m_2$, two cubic vertices are given in \rf{03032020-man-45}, \rf{03032020-man-46}.

As a remark, we note that all vertices \rf{03032020-man-43}-\rf{03032020-man-46} can be presented on an equal footing as
\beq
\label{03032020-man-46-a1} && \hspace{-0.9cm} p_\smpth^- = \frac{C_{\epsilon_1\epsilon_2}}{\beta_3} \big(1 + \frac{\irm \epsilon_2 \Po}{m_2 \beta_3} \big) \big(\frac{\beta_3}{\beta_1}\big)^{\epsilon_1 (s_1+\half) + \half}
\big(\frac{\beta_3}{\beta_2}\big)^{\epsilon_2 (s_2+\half) +\half}\,, \hspace{1cm} \epsilon_1 m_1 + \epsilon_2 m_2 =0 \,, \qquad
\\
\label{03032020-man-46-a2}&&  \hspace{-0.9cm} C_{-11} = C_+\,, \qquad C_{1-1} = \Cb_+\,, \qquad C_{-1-1} = C_-\,, \qquad
C_{11} = \Cb_-\,,
\eeq
where $\epsilon_1^2=1$, $\epsilon_2^2=1$. In \rf{03032020-man-46-a2}, we match coupling constants in \rf{03032020-man-43}-\rf{03032020-man-46} and \rf{03032020-man-46-a1}.

\noindent {\bf Higher-derivative form}. For the vertices in \rf{03032020-man-43},\rf{03032020-man-44}, and \rf{03032020-man-46}, we find the following higher-derivative representation,
{\small
\beq
\label{03032020-man-47} && \hspace{-2cm} \hbox{\small\bf IIb1.1)} \hspace{0.6cm} p_\smpth^- =  C_+^\HDsm   \frac{1}{\beta_2} (\frac{\beta_3}{\beta_1})^{-s_1}  L_2^{s_2}\,, \hspace{0.6cm}  C_+^\HDsm = \frac{2 C_+}{(2m_2)^{s_2}}\,, \hspace{2.3cm} m_1=m_2\,,
\\
\label{03032020-man-48} && \hspace{-2cm} \hbox{\small\bf IIb1.2)} \hspace{0.6cm} p_\smpth^-  = \Cb_+^\HDsm \frac{1}{\beta_1} L_1^{s_1}  (\frac{\beta_3}{\beta_2})^{-s_2}\,, \hspace{0.6cm}  \Cb_+^\HDsm = \frac{2 \Cb_+}{(-2m_2)^{s_1}}\,, \hspace{2cm} m_1=m_2\,,
\\
\label{03032020-man-49} && \hspace{-2cm} \hbox{\small\bf IIb1.4)} \hspace{0.6cm} p_\smpth^- =  \Cb_-^\HDsm  \frac{\beta_3}{\beta_1\beta_2} L_1^{s_1}  L_2^{s_2}\,, \hspace{0.9cm}  \Cb_-^\HDsm = \frac{2 \Cb_-}{(2m_2)^{s_1+s_2}}\,, \hspace{1.8cm} m_1=-m_2\,,
\\
\label{03032020-man-50} && \hspace{2cm} L_1 \equiv     \frac{1}{\beta_1}\big(\irm \Po - m_1 \beta_3\big)\,,
\hspace{1cm} L_2 \equiv     \frac{1}{\beta_2}\big(\irm \Po + m_2 \beta_3\big)\,,
\eeq
}
where $C_+^\HDsm$, $\Cb_\pm^\HDsm$ are coupling constants entering higher-derivative form of cubic vertices. In \rf{03032020-man-47}-\rf{03032020-man-49}, we show how these constants are related to the coupling constants entering first-derivative form of cubic vertices \rf{03032020-man-43}-\rf{03032020-man-46}.  Expressions for the  $j_\smpth^{-1}$ corresponding to $p_\smpth^-$ in \rf{03032020-man-47}-\rf{03032020-man-49} are given by
{\small
\beq
\label{03032020-man-47-j1} && \hspace{-2cm} \hbox{\small\bf IIb1.1)} \hspace{0.6cm} j_\smpth^{-1} = C_+^\HDsm \Big( 2\irm (s_1 - \frac{\beta_3}{2\beta_2} \big) -  \frac{2\irm \betach_2}{3\beta_2}s_2\Big) \frac{1}{\beta_2}(\frac{\beta_3}{\beta_1})^{-s_1} L_2^{s_2-1}\,,
\hspace{0.7cm} m_1=m_2\,, \hspace{0.5cm} s_2\geq 1;
\\
\label{03032020-man-48-j2} && \hspace{-2cm} \hbox{\small\bf IIb1.2)} \hspace{0.6cm} j_\smpth^{-1} = - \Cb_+^\HDsm \Big( 2\irm (s_2 - \frac{\beta_3}{2\beta_1} \big) + \frac{2\irm \betach_1}{3\beta_1}s_1\Big) \frac{1}{\beta_1} L_1^{s_1 - 1}  (\frac{\beta_3}{\beta_2})^{-s_2}\,,
\hspace{0.4cm} m_1=m_2\,, \hspace{0.5cm} s_1\geq 1;
\\
\label{03032020-man-49-j3} && \hspace{-2cm} \hbox{\small\bf IIb1.4)} \hspace{0.6cm} j_\smpth^{-1} =   \frac{\irm \betach_3\beta_3 \Cb_-^\HDsm }{\beta_1^2\beta_2^2}\big( \irm \Po - m_1 \beta_3\big) L_1^{s_1-1} L_2^{s_2-1}  -  \sum_{a=1,2} \frac{2\irm \betach_a}{3\beta_a} \partial_{L_a}p_\smpth^- \,, \hspace{0.8cm} m_1=-m_2\,,
\eeq
}
where, in \rf{03032020-man-49-j3}, $s_1+s_2\geq 1$. Note that, in \rf{03032020-man-49-j3}, $\beta_2 L_1 = \beta_1 L_2$. We see that first-derivative vertices \rf{03032020-man-43}-\rf{03032020-man-46} are valid for arbitrary $s_1$, $s_2$ in \rf{03032020-man-42}, while, for the higher-derivative vertices, we should use restrictions on $s_1$, $s_2$ in \rf{03032020-man-47-j1},\rf{03032020-man-48-j2} and the restriction $s_1+s_2\geq 1$ in \rf{03032020-man-49-j3}.

As a remark, we note that, by using $L_{\crit, a}$ \rf{02032020-man-24-c3}, we find alternative higher-derivative form for some first-derivative vertices \rf{03032020-man-43}-\rf{03032020-man-46} given by
\beq
\label{03032020-man-49-j4} &&  p_\smpth^- =  C_{\epsilon_1\epsilon_2}^\HDsm (\irm \Po +\Po_{\epsilon m})^{\Sbf_\epsilon^\f} \beta_1^{-\epsilon_1 (s_1+\half) - \half} \beta_2^{-\epsilon_2 (s_2+\half) -\half} \,, \hspace{1cm}  C_{\epsilon_1\epsilon_2}^\HDsm =  2 (2\epsilon_2 m_2)^{-\Sbf_\epsilon^\f} C_{\epsilon_1\epsilon_2}\,, \qquad
\\
\label{03032020-man-49-j5} &&\hspace{1cm}  \Po_{\epsilon m} = \epsilon_2 m_2 \beta_3\,, \hspace{0.5cm} \Sbf_\epsilon^\f = \epsilon_1 (s_1+\half) + \epsilon_2 (s_2 +\half)\,, \hspace{0.5cm} \epsilon_1^2=\epsilon_2^2=1\,.
\eeq
Vertices \rf{03032020-man-49-j4} and corresponding $j_\smpth^{-1}$ are obtained by setting $m_3=0$, $s_3=0$ in \rf{03032020-man-23-b1} and \rf{03032020-man-23-b3} respectively. This gives the restriction $\Sbf_\epsilon^\f > 0$ in \rf{03032020-man-49-j4} which is redundant  for first-derivative vertices \rf{03032020-man-43}-\rf{03032020-man-46}. Thus, higher-derivative vertices \rf{03032020-man-49-j4} provide us the particular vertices, while first-derivative vertices \rf{03032020-man-43}-\rf{03032020-man-46} provide us the full list of vertices.

\noindent {\bf IIb2)  Cubic vertices for one fermionic massless field, one fermionic arbitrary spin massive fields and one bosonic arbitrary spin massive field with masses $|m_2|=|m_3|$}. Using notation as in \rf{27022020-man-08-a1}-\rf{27022020-man-08-a6}, we consider a cubic vertex for three fields with the following masses and spins:
\beq
\label{03032020-man-51} && (m_1,\half)-(m_2,s_2+\half)-(m_3,s_3)\,, \hspace{1.2cm} s_2,s_3 \in \No_0\,,
\nonumber\\
\label{03032020-man-52} &&  m_1 = 0\,, \quad m_2\ne 0\,, \quad m_3 \ne 0 \quad |m_2|=|m_3|\,.
\eeq
The constraint $|m_2|=|m_3|$ leads to the consideration of two different cases: $m_2=m_3$ and $m_2=-m_3$. For these two cases, we find four vertices which we label to as {\small\bf IIb2.1-IIb2.4}. First-derivative form of these four vertices is given by
\beq
\label{03032020-man-53} && \hbox{\small\bf IIb2.1)} \hspace{1.5cm} p_\smpth^- =  C_+ \frac{1}{\beta_1} (1 + \frac{\irm \Po}{m_3 \beta_1 }) (\frac{\beta_1}{\beta_2})^{-s_2}  (\frac{\beta_1}{\beta_3})^{s_3}\,,
\hspace{1.5cm}  m_2=m_3; \hspace{3cm}
\\
\label{03032020-man-54} && \hbox{\small\bf IIb2.2)} \hspace{1.5cm} p_\smpth^-  = \Cb_+   \frac{1}{\beta_2} (1 - \frac{\irm \Po}{m_3 \beta_1} ) (\frac{\beta_1}{\beta_2})^{s_2}  (\frac{\beta_1}{\beta_3})^{-s_3}\,, \hspace{1.5cm}  m_2=m_3;
\\
\label{03032020-man-55} && \hbox{\small\bf IIb2.3)} \hspace{1.5cm}  p_\smpth^- =  C_- \frac{1}{\beta_2} (1 + \frac{\irm \Po}{m_3 \beta_1 }) (\frac{\beta_1}{\beta_2})^{s_2}  (\frac{\beta_1}{\beta_3})^{s_3}\,,
\hspace{1.8cm} m_2 = - m_3;
\\
\label{03032020-man-56} && \hbox{\small\bf IIb2.4)} \hspace{1.5cm}  p_\smpth^-  = \Cb_-  \frac{1}{\beta_1} (1 - \frac{\irm \Po}{m_3 \beta_1} ) (\frac{\beta_1}{\beta_2})^{-s_2}  (\frac{\beta_1}{\beta_3})^{-s_3}\,,
\hspace{1.3cm} m_2= - m_3\,.
\eeq
Thus, for masses $m_2=m_3$, we find two cubic vertices given in \rf{03032020-man-53}, \rf{03032020-man-54}, while,
for masses $m_2=-m_3$, two cubic vertices are given in \rf{03032020-man-55}, \rf{03032020-man-56}.

As a remark, we note that all vertices \rf{03032020-man-53}-\rf{03032020-man-56} can be presented on an equal footing as
\beq
\label{03032020-man-56-a1} && \hspace{-0.5cm} p_\smpth^- = \frac{C_{\epsilon_2\epsilon_3}}{\beta_1} \big(1 + \frac{\irm \epsilon_3 \Po}{m_3 \beta_1} \big) \big(\frac{\beta_1}{\beta_2}\big)^{\epsilon_2 (s_2+\half)  +\half }
\big(\frac{\beta_1}{\beta_3}\big)^{\epsilon_3 s_3}\,, \hspace{1cm} \epsilon_2 m_2 + \epsilon_3 m_3 =0 \,, \qquad
\\
\label{03032020-man-56-a2}&&  \hspace{-0.5cm} C_{-11} = C_+\,, \qquad C_{1-1} = \Cb_+\,, \qquad C_{11} = C_-\,, \qquad
C_{-1-1} = \Cb_-\,,
\eeq
where, $\epsilon_2^2=\epsilon_3^2=1$, and, in \rf{03032020-man-56-a2}, we identify coupling constants in \rf{03032020-man-53}-\rf{03032020-man-56} and \rf{03032020-man-56-a1}.

\noindent {\bf Higher-derivative form}. For the vertices in \rf{03032020-man-53},\rf{03032020-man-54}, and \rf{03032020-man-56}, we find the following higher-derivative representation,
{\small
\beq
\label{03032020-man-57} && \hspace{-2cm} \hbox{\small\bf IIb2.1)} \hspace{0.6cm} p_\smpth^- =  C_+^\HDsm    \frac{1}{\beta_1} (\frac{\beta_1}{\beta_2})^{-s_2}  L_3^{s_3}\,, \hspace{1cm}  C_+^\HDsm = \frac{2C_+}{(2m_3)^{s_3}} \,, \hspace{2.2cm} m_2=m_3\,;
\\
\label{03032020-man-58} && \hspace{-2cm} \hbox{\small\bf IIb2.2)} \hspace{0.6cm} p_\smpth^-  = \Cb_+^\HDsm   \frac{1}{\beta_2} L_2^{s_2}  (\frac{\beta_1}{\beta_3})^{-s_3}\,, \hspace{1cm}  \Cb_+^\HDsm = \frac{2\Cb_+}{(-2m_3)^{s_2}} \,, \hspace{1.9cm} m_2=m_3\,;
\\
\label{03032020-man-59} &&  \hspace{-2cm} \hbox{\small\bf IIb2.3)} \hspace{0.6cm}  p_\smpth^- =  \Cb_-^\HDsm  \frac{1}{\beta_2} L_2^{s_2}  L_3^{s_3}\,, \hspace{1.7cm}  \Cb_-^\HDsm = \frac{2\Cb_-}{(2m_3)^{s_2+s_3}} \,, \hspace{1.7cm} m_2=-m_3\,;
\\
\label{03032020-man-60} && L_2 \equiv    \frac{1}{\beta_2}\big(\irm \Po - m_2 \beta_1\big)\,, \hspace{1cm}  L_3 \equiv     \frac{1}{\beta_3}\big(\irm \Po + m_3 \beta_1\big)\,,
\eeq
}
where $C_+^\HDsm$, $\Cb_\pm^\HDsm$ are coupling constants of higher-derivative cubic vertices. In \rf{03032020-man-57}-\rf{03032020-man-59}, we show how these constants are related to the coupling constants entering first-derivative form of cubic vertices \rf{03032020-man-53}-\rf{03032020-man-56}.
Expressions for $j_\smpth^{-1}$ corresponding to $p_\smpth^-$ in \rf{03032020-man-57}-\rf{03032020-man-59} are given by
{\small
\beq
\label{03032020-man-57-j1} && \hspace{-2cm} \hbox{\small\bf IIb2.1)} \hspace{0.6cm} j_\smpth^{-1} =   C_+^\HDsm \Big(  2\irm (s_2+\half)   -    \frac{2\irm \betach_3}{3\beta_3} s_3 \Big) \frac{1}{\beta_1}(\frac{\beta_1}{\beta_2})^{-s_2}   L_3^{s_3-1}\,, \hspace{1cm} m_2=m_3\,, \hspace{0.5cm} s_3\geq 1;
\\
\label{03032020-man-58-j2} && \hspace{-2cm} \hbox{\small\bf IIb2.2)} \hspace{0.6cm} j_\smpth^{-1} =  - \Cb_+^\HDsm \Big(  2\irm (s_3 + \frac{\beta_3}{2\beta_2})   + \frac{2\irm \betach_2}{3\beta_2} s_2 \Big) \frac{1}{\beta_2} L_2^{s_2-1}  (\frac{\beta_1}{\beta_3})^{-s_3}\,,
\hspace{0.4cm} m_2=m_3\,, \hspace{0.5cm} s_2\geq 1;
\\
\label{03032020-man-59-j3} &&  \hspace{-2cm} \hbox{\small\bf IIb2.3)} \hspace{0.6cm}  j_\smpth^{-1} =  - \frac{\irm \Cb_-^\HDsm}{\beta_2^2}\big( \irm \Po - m_2 \beta_1\big) L_2^{s_2-1} L_3^{s_3-1}  -    \sum_{a=2,3} \frac{2\irm \betach_a}{3\beta_a} \partial_{L_a}p_\smpth^- \,,
\hspace{1cm} m_2=-m_3\,;
\eeq
}
where, in \rf{03032020-man-59-j3}, we assume the restriction $s_2+s_3\geq 1$.
Note that, in \rf{03032020-man-59-j3}, $\beta_3 L_2 = \beta_2 L_3$.

As a remark, we note that, by using $L_{\crit, a}$ \rf{02032020-man-24-c3}, we find alternative higher-derivative form for some first-derivative vertices \rf{03032020-man-53}-\rf{03032020-man-56} given by
\beq
\label{03032020-man-59-j4} && \hspace{-0.5cm} p_\smpth^- =  C_{\epsilon_2\epsilon_3}^\HDsm (\irm \Po +\Po_{\epsilon m})^{\Sbf_\epsilon^\f} \beta_1^{-\half \epsilon_1  - \half} \beta_2^{-\epsilon_2 (s_2+\half) -\half} \beta_3^{-\epsilon_3 s_3 }\,,   \hspace{1cm}  C_{\epsilon_2\epsilon_3}^\HDsm =  2 (2\epsilon_3 m_3)^{-\Sbf_\epsilon^\f} C_{\epsilon_2\epsilon_3}\,, \qquad\quad
\\
\label{03032020-man-59-j5} && \hspace{0.5cm} \Po_{\epsilon m} = \epsilon_3 m_3 \beta_1, \hspace{1cm} \Sbf_\epsilon^\f = \half \epsilon_1 + \epsilon_2 (s_2+\half) + \epsilon_3 s_3, \qquad \epsilon_1^2=\epsilon_2^2=\epsilon_3^2=1\,.
\eeq
Vertices $p_\smpth^-$ \rf{03032020-man-59-j4} and corresponding $j_\smpth^{-1}$ are obtained by setting $m_1=0$, $s_1=0$ in \rf{03032020-man-23-b1} and \rf{03032020-man-23-b3} respectively. This implies the restriction $\Sbf_\epsilon^\f > 0$ in \rf{03032020-man-59-j4} which is redundant for first-derivative vertices \rf{03032020-man-53}-\rf{03032020-man-56}. Thus, higher-derivative vertices \rf{03032020-man-59-j4} provide us the particular list of vertices, while   first-derivative vertices \rf{03032020-man-53}-\rf{03032020-man-56} provide us the full list of vertices.

\noindent {\bf{\small  IIIa}) Cubic  vertex for two fermionic massless fields and one arbitrary spin massive field}. Using notation as in \rf{27022020-man-08-a1}-\rf{27022020-man-08-a6}, we consider a cubic vertex for three fields with the following masses and spins:
\be \label{03032020-man-61}
(0,\half)-(0,\half)-(m_3,s_3)\,, \hspace{1cm} m_3 \ne 0 \hspace{1cm} s_3 \in \No_0\,.
\ee
Solution for vertices $V$, $\Vb$ entering cubic vertex $p_\smpth^-$ \rf{01032020-man-x01} is given by
\beq
\label{03032020-man-63} && \hspace{-1cm} V = C V_\kappa\,, \qquad \Vb = C V_{-\kappa}\,,
\\
\label{03032020-man-64} && V_\kappa \equiv \frac{1}{\kappa} L_{\kappa,3}^{s_3}\,,
\hspace{1cm}  L_{\kappa,3} \equiv  \frac{ \kappa}{\beta_3} + \frac{\betach_3}{2\beta_3} m_3   \,, \hspace{1cm} \kappa^2 = -\beta_1\beta_2 m_3^2\,,
\eeq
where $C$ is coupling constant. Plugging $V$ and $\Vb$ \rf{03032020-man-63} into \rf{01032020-man-x01}, we get first-derivative form of the cubic vertex $p_\smpth^-$.

\noindent {\bf Higher-derivative form}. Higher-derivative form of the cubic vertex in \rf{03032020-man-61} is obtained by using expressions for $V$, $\Vb$ \rf{03032020-man-63} and $p_\smpth^-$ \rf{01032020-man-x01} (for some details, see Appendix A). Doing so, we get
\be
\label{03032020-man-66}   p_\smpth^- = C^\HDsm K L_3^{s_3}\,, \hspace{0.7cm} K \equiv \frac{\irm \Po}{\beta_1\beta_2}\,, \hspace{0.7cm} L_3 \equiv  \irm \frac{  \Po}{\beta_3} + \frac{\betach_3}{2\beta_3} m_3   \,,  \hspace{1cm} C^\HDsm = - \frac{1}{m_3^2} C \,,
\ee
where the coupling constant $C$ appearing in \rf{03032020-man-66} is the same as the one in \rf{03032020-man-63}.
Expression for $j_\smpth^{-1}$ can be obtained by using    $p_\smpth^-$ \rf{03032020-man-66} and general relations in \rf{01032020-man-12},\rf{01032020-man-14},\rf{01032020-man-18}.

\noindent {\bf{\small  IIIb}) Cubic  vertex for one fermionic arbitrary spin massive field, one fermionic massless field, and one scalar massless field}. Using notation as in \rf{27022020-man-08-a1}-\rf{27022020-man-08-a6}, we consider a cubic vertex for three fields with the following masses and spins:
\be \label{03032020-man-68}
(m_1,s_1+\half)-(0,\half)-(0,0)\,, \hspace{1cm}  m_1 \ne 0,\hspace{1cm} s_1 \in \No_0\,.
\ee
Solution for vertices $V$, $\Vb$ entering cubic vertex $p_\smpth^-$ \rf{01032020-man-x01} is given by
\beq
\label{03032020-man-70} && V = C V_\kappa\,, \qquad \Vb = C V_{-\kappa}\,,\qquad V_\kappa \equiv K_\kappa L_{\kappa,1}^{s_1}\,,
\\
\label{03032020-man-71} &&    K_\kappa \equiv \frac{1}{\beta_1\beta_2}(\kappa + m_1\beta_2)\,,
\qquad  L_{\kappa,1} \equiv  \frac{ \kappa}{\beta_1} + \frac{\betach_1}{2\beta_1} m_1   \,, \hspace{0.4cm} \kappa^2 = - \beta_2 \beta_3 m_1^2\,,\qquad
\eeq
where $C$ is coupling constant. Plugging $V$ and $\Vb$ \rf{03032020-man-70} into \rf{01032020-man-x01}, we get first-derivative form of the cubic vertex $p_\smpth^-$.

\noindent {\bf Higher-derivative form}. Higher-derivative form of the cubic vertex in \rf{03032020-man-68} is obtained by plugging $\kappa=\irm \Po$ into expressions for $V$, $\Vb$ \rf{03032020-man-70} and $p_\smpth^-$ \rf{01032020-man-x01}. Doing so, we get
\be \label{03032020-man-73}
p_\smpth^- =   C K L_1^{s_1}\,, \hspace{1cm} K \equiv \frac{1}{\beta_1\beta_2}(\irm \Po + m_1\beta_2)\,, \hspace{1cm} L_1 \equiv  \irm \frac{ \Po}{\beta_1} + \frac{\betach_1}{2\beta_1} m_1  \,,
\ee
where the coupling constant $C$ in \rf{03032020-man-73} is the same as the one in \rf{03032020-man-70}.
Expression for $j_\smpth^{-1}$ can be obtained by using    $p_\smpth^-$ \rf{03032020-man-73} and general relations in \rf{01032020-man-12},\rf{01032020-man-14},\rf{01032020-man-18}.

\noindent {\bf IV) Cubic  vertices for two fermionic massless fields and one scalar massless field}. Using notation as in \rf{27022020-man-08-a1}-\rf{27022020-man-08-a6}, we consider a cubic vertex for two fermionic massless  fields and one massless scalar field:
\be \label{03032020-man-75}
(0,\half)-(0,\half)-(0,0)\,.
\ee
For this case $\kappa=0$ and we use basis of cubic vertices $V_0$, $V_1$ with  equations \rf{01032020-man-09} and equations obtained by setting $m_1=0$, $m_2=0$, $m_3=0$ in \rf{01032020-man-06},\rf{01032020-man-07}. For vertices $V_0$ and $V_1$, we get equations \rf{01032020-man-09} and equation $\No_\beta^E V_0=0$. The $\beta$-analytic solution to these equations is given by
\be \label{03032020-man-77}
V_0 = 0\,, \hspace{1cm} V_1 = \frac{1}{\beta_1\beta_2} v\big( \frac{\beta_1}{\beta_2}\big)\,,
\ee
where $v$ is $\beta$-analytic.  As before for bose vertices, we see that, for three massless fields, we have an infinite number of light-cone gauge vertices \rf{03032020-man-77} parametrized by  the function $v$.  For $v=const$, the vertex $V_1$ is associated with the $\bar\psi\psi \phi$ Yukawa interaction in Lorentz covariant approach. For $v=\betach_3/\beta_3$, i.e.,  $v=1-\frac{\beta_1}{\beta_2}/1+\frac{\beta_1}{\beta_2}$, the vertex $V_1$ \rf{03032020-man-77} is associated with the cubic vertex of Yang-Mills theory.

%%%%%%%%%%%%%%%%%%%%%%%%%%%%%%%%%%%%%%%%%%%%%%%%%%%%%%%%%%%%%%%%%%
\newsection{ \large Conclusions}\label{fin-concl}
%%%%%%%%%%%%%%%%%%%%%%%%%%%%%%%%%%%%%%%%%%%%%%%%%%%%%%%%%%%%%%%%%%

In this paper, we exploited light-cone gauge approach for studying interacting arbitrary integer and half-integer spin massive fields and scalar and one-half massless spin fields propagating flat $3d$ space. For such fields we build all cubic interaction vertices. As it is happens in any other approaches, the light-cone gauge interaction vertices have freedom related to field redefinitions. Using this freedom we worked out two equivalent forms of light-cone gauge vertices which we refer to as first-derivative form and higher-derivative form. We expect that our results could have a lot of very interesting generalizations and applications which we now discuss.

\medskip
\noindent \ibf) In this paper, we built cubic interactions vertices for both the bosonic and fermionic massive (massless) fields in $R^{2,1}$. We expect therefore that our results for cubic interaction vertices provide good starting point for studying supersymmetric interacting massive (massless) fields in $R^{2,1}$. Light-cone gauge approach turns out to be convenient for studying massless higher-spin theories in $R^{3,1}$. Using this approach, all cubic interaction vertices for massless arbitrary spin $\NN$-extended supermultiplets in $R^{3,1}$ have been found in Refs.\cite{Metsaev:2019aig,Metsaev:2019dqt}%
\footnote{ Discussion of light-cone gauge cubic interactions for scalar $\NN$-extended supermultiplet in $R^{3,1}$ may be found in Ref.\cite{Bengtsson:1983pg} (see also recent discussion in Ref.\cite{Ananth:2020mws}).
}
As the light-cone gauge formulation of supersymmetric massive fields in $R^{2,1}$ is similar to the light-cone gauge formulation of massless fields in $R^{3,1}$ we think that studying of supersymmetric massive field theory in $R^{2,1}$ by using light-cone gauge approach will be fruitful. In the framework of Lorentz covariant approach, the study of free massive fields in $3d$ may be found in Refs.\cite{Zinoviev:2015sra}-\cite{Buchbinder:2014ata}, while the supersymmetric free massive field theories are investigated in Refs.\cite{Buchbinder:2017izy}-\cite{Buchbinder:2016jgk}.%
\footnote{ For study of massless higher-spin supermuliplets in $3d$, see, e.g., Refs.\cite{Hutomo:2018iqo}. Discussion of reducible massless higher-spin supermultiplets may be found in Ref.\cite{Sorokin:2018djm}.}
The gravitational interaction of higher-spin massive fields in $3d$ is considered in Ref.\cite{Buchbinder:2012xa}.

\noindent \iibf) In the above-presented study, we restricted our attention to interacting arbitrary spin massive and low spin massless fields propagating in $R^{2,1}$. It would be very interesting to generalize our study to massive (massless) fields propagating in $AdS_3$. Such generalization would be important in view of potentially interesting applications to superstring theory in $AdS_3$ space because one expects that it is massive arbitrary spin fields and low spin massless fields that enter spectrum of states of a superstring in AdS background. Light-cone gauge formulation of free arbitrary spin massive fields propagating in $AdS_3$ space was obtained long ago in Ref.\cite{Metsaev:1999ui,Metsaev:2000qb}, while
method for studying interacting light-cone fields in AdS space was developed in Ref.\cite{Metsaev:2018xip}.%
\footnote{ In Ref. \cite{Metsaev:2018xip}, cubic vertices for massless arbitrary spin light-cone gauge fields in $AdS_4$ were obtained. Interesting applications of cubic vertices of massless arbitrary spin light-cone gauge fields in $AdS_4$ for studying $3d$ Chern-Simons matter theories via the $AdS/CFT$ correspondence may be found in Ref.\cite{Skvortsov:2018uru}.}
We expect therefore that method and results in Refs.\cite{Metsaev:2000qb,Metsaev:2018xip} will alow us to study interacting massive fields in $AdS_3$. We note also, that results for cubic vertices of massless fields in $AdS_4$ in Ref.\cite{Metsaev:2018xip} and procedure of dimensions reduction (digression) in AdS space developed in Refs.\cite{Metsaev:2000qb,Artsukevich:2008vy} open interesting possibility for studying cubic vertices of massive fields in $AdS_3$. We think also that knowledge of light-cone gauge formulation of interacting massive fields in $AdS_3$ will give new opportunities for study of $AdS/CFT$ correspondence along the lines in Refs.\cite{Beccaria:2019stp}-\cite{Alkalaev:2019xuv}.

\noindent \iiibf) Conformal higher-spin theories in $3d$ space have attracted some interest in the recent time (see, e.g., Refs.\cite{Nilsson:2013tva}-\cite{Grigoriev:2019xmp}).
In this respect we note that the light-cone gauge formulation of ordinary-derivative free and interacting totally symmetric conformal fields propagating in $R^{d-1,1}$ space with even $d$ was developed in Ref.\cite{Metsaev:2016rpa}, while the ordinary-derivative Lorentz covariant formulation of conformal fields in $3d$ space is available from Ref.\cite{Metsaev:2016oic}.%
\footnote{ We recall that, in higher dimensions, besides totally symmetric conformal fields, there exist so called mixed-symmetry conformal fields (see, e.g., Refs.\cite{Vasiliev:2009ck}).}
Method in Ref.\cite{Metsaev:2016rpa} provides possibility to re-cast the Lorentz covariant ordinary-derivative formulation in Ref.\cite{Metsaev:2016oic} into the light-cone gauge formulation. As the light-cone gauge formulation considerably simplifies the whole analysis we expect then that the light-cone gauge formulation of conformal fields in $3d$ space will be useful for better understanding conformal higher-spin fields in $3d$ space.

\noindent \ivbf) Recently, the light-cone gauge approach has fruitfully been used for studying loop corrections in higher-spin theories. In Refs.\cite{Skvortsov:2018jea,Skvortsov:2020gpn}, it was shown that the massless
higher-spin chiral theory \cite{Ponomarev:2016lrm} is free of
one-loop divergencies and some arguments were given for cancellation
of all loop divergencies. We note that, as discussed in
Refs.\cite{Fradkin:1982kf,Beccaria:2014xda},  upon use of a proper
regularization scheme, the quantum properties of some low spin
supersymmetric field  theories and their dimensionally reduced
counterparts
(with all massive  KK modes included)  are equivalent. In this
respect,  it would be interesting to investigate quantum properties of
massless higher-spin chiral theory by using its dimensionally reduced
counterpart which is realized as a massive higher-spin chiral theory
in $3d$. For the description of massive higher-spin chiral theory in
$3d$, see Appendix A.

\noindent \vbf) As we noted, the light-cone formulations of massive fields in $R^{2,1}$ and massless fields in $R^{3,1}$ share some features. The light-cone gauge approach was used in Refs.\cite{Metsaev:1991mt,Metsaev:1991nb} for studying quartic interaction vertices of massless fields in $4d$. We expect then that some considerations in Refs.\cite{Metsaev:1991mt,Metsaev:1991nb} can in relatively straightforward way be generalized to quartic vertices for massive fields in $3d$. We think also that study of quartic vertices for massive fields along the lines in Refs.\cite{Hinterbichler:2017qcl}  could be of some interest. Interesting recent discussion of $n$-point vertices of higher-spin fields may be found in Refs.\cite{Joung:2019wbl}. See also general discussion by using light-cone gauge approach in Ref.\cite{Ponomarev:2016cwi}.

\noindent \vibf) For analysis of various hupergravity  theories in $3d$, BRST technique was used in Ref.\cite{Rahman:2019mra}, while in Ref.\cite{Buchbinder:2015kca} BRST approach was used for studying massless fields in $4d$. Note that that it is use of twistor-like variables that considerably simplifies BRST analysis in Ref.\cite{Buchbinder:2015kca}.  Interesting use of twistor-like variables for studying string model in AdS space may be found in Refs.\cite{Uvarov:2016slb}.
Extension of ideas and approaches in Refs.\cite{Rahman:2019mra}-\cite{Uvarov:2016slb} to the case of massive fields in $4d$ could of some interest.

\noindent \viibf) In this paper, we studied integer and half-integer spin fields. Poincar\'e  algebra admits continuous-spin fields. Cubic vertices for light-cone gauge continuous-spin massive and massless fields in $R^{d-1,1}$ with $d\geq 4$ were studied in Refs.\cite{Metsaev:2017cuz,Metsaev:2018moa}. We think that extension of our studies to the case of continuous spin field could be very interesting. For the reader convenience, we note that
Lorentz covariant cubic vertex for coupling of massless and massive continuous-spin field to two massive scalar fields
were studied in Refs.\cite{Bekaert:2017xin}. BRST approach for studying continuous-spin free field was applied in Refs.\cite{Bengtsson:2013vra,Metsaev:2018lth,Buchbinder:2018yoo}, while various Lagrangian formulations of supersymmetric continuous-spin free field were considered in Refs.\cite{Zinoviev:2017rnj}. For AdS space, light-cone gauge Lagrangian for continuous-spin free field was obtained in Refs.\cite{Metsaev:2017myp,Metsaev:2019opn}, while metric-like and frame-like covariant Lagrangian formulations were studied in the respective Refs.\cite{Metsaev:2016lhs,Metsaev:2017ytk} and Ref.\cite{Khabarov:2017lth}.

\medskip

{\bf Acknowledgments}. This work was supported by the RFBR Grant No.20-02-00193.

%%%%%%%%%%%%%%%%%%%%%%%%%%%%%%%%%%%%%%%%%%%%%%%%%%%%%%%%%%%%%%%%%%%%%%%%%%%%%%%%%%%%%%%%%%%%%%%%%%%%%%%
%%%%%%%%%%%%%%%%%%%%%%%%%%%%%%%%%%%%%%%%%%%%%%%%%%%%%%%%%%%%%%%%%%%%%%%%%%%%%%%%%%%%%%%%%%%%%%%%%%%%%%%
\setcounter{section}{0}\setcounter{subsection}{0}
\appendix{ \large Derivation of cubic vertices }
%%%%%%%%%%%%%%%%%%%%%%%%%%%%%%%%%%%%%%%%%%%%%%%%%%%%%%%%%%%%%%%%%%%%%%%%%
%%%%%%%%%%%%%%%%%%%%%%%%%%%%%%%%%%%%%%%%%%%%%%%%%%%%%%%%%%%%%%%%%%%%%%%%%

\noindent {\bf Properties of $L_{\kappa, a}$}. Quantities $L_{\kappa, a}$  \rf{02032020-man-13} and $L_a$ \rf{01032020-man-15}, $L_{\crit,a}$, $\Lb_{\crit,a}$ \rf{02032020-man-24-c3} are building blocks for our first-derivative and higher-derivative cubic vertices respectively. Basic properties of $L_{\kappa, a}$, $L_a$, and  $L_{\crit,a}$ are given by the following differential relations
\beq
\label{21052020-man-01} && \frac{\kappa}{\beta} \No_\beta L_{\kappa,a}  =  \frac{m_a}{\beta_a}  L_{\kappa,a}\,,
\\
\label{21052020-man-04} && \Big( -\frac{\Po}{\beta} \No_\beta  + \sum_{b=1,2,3}  \frac{\betach_b }{6\beta_b} m_b^2 \partial_{\Po} \Big) L_a = \frac{2\irm \betach_a}{3\beta_a} \Pbf^- + \frac{\irm m_a}{\beta_a} L_a\,,
\\
&& \Big( -\frac{\Po}{\beta} \No_\beta  + \sum_{b=1,2,3}  \frac{\betach_b }{6\beta_b} m_b^2 \partial_{\Po} \Big) L_{\crit,a} = \frac{2\irm \betach_a}{3\beta_a}  \Pbf^- + \frac{\irm \epsilon_a m_a}{\beta_a} L_{\crit,a}\,,
\eeq
where we use notation in \rf{29022020-man-03}, \rf{29022020-man-11}-\rf{29022020-man-13}.
The $L_{\crit,a}$ is obtained from $L_a$ by the replacements $m_b\rightarrow \epsilon_b m_b$ for all $b=1,2,3$ and by using the constraint $\Pbf_{\epsilon m}=0$.  The  $\Lb_{\crit,a}$ is obtained from $L_{\crit,a}$ by the replacements $\epsilon_b\rightarrow -\epsilon_b$ for all $b=1,2,3$. We note the algebraic relations
\beq
&& L_{\kappa,a} L_{-\kappa,a}  = \frac{1}{4m_a^2} D \,,
\\
\label{21052020-man-02} && L_{\kappa,1} L_{\kappa,2} = \frac{m_{231}m_{312}}{4\beta_1\beta_2 m_1 m_2} K_{12}^2\,,
\\
&& L_{\kappa,1} L_{\kappa,2}L_{\kappa,3} = - \frac{m_{123}m_{231}m_{312}}{8\beta m_1 m_2 m_3} K_{12}K_{23}K_{31}\,,
\\
\label{21052020-man-03}&& K_{ab} \equiv \kappa + m_a \beta_b - m_b \beta_a\,, \hspace{1cm}
m_{abc} \equiv m_a + m_b - m_c\,,
\\
&& L_{\crit, a} \Lb_{\crit, a} = - \frac{2\beta}{\beta_a^2} \Pbf^-\,,
\eeq
where $\Pbf^-$, $\beta$, and $D$ are given in \rf{29022020-man-11}, \rf{02032020-man-01}.

We now outline the derivation of our cubic vertices in turn.

\noindent {\bf Cubic vertex \rf{02032020-man-12}}. In place of vertices $V$, $\Vb$, we introduce new vertices $V^{(1)}$, $\Vb^{(1)}$,
\be \label{21052020-man-15}
V =   L_{\kappa,1}^{s_1} L_{\kappa,2}^{s_2} L_{\kappa,3}^{s_3} V^{(1)}\,, \hspace{1cm}
\Vb =   L_{-\kappa,1}^{s_1} L_{-\kappa,2}^{s_2} L_{-\kappa,3}^{s_3} \Vb^{(1)}\,.
\ee
In terms of new vertices, equations for $V$, $\Vb$  \rf{01032020-man-x03},\rf{01032020-man-x04} take the form
\be \label{21052020-man-16}
\No_\beta V^{(1)} = 0 \,, \qquad \No_\beta \Vb^{(1)} = 0 \,, \qquad  \sum_{a=1,2,3}  \beta_a\partial_{\beta_a} V^{(1)} =0 \,. \qquad  \sum_{a=1,2,3}  \beta_a\partial_{\beta_a} \Vb^{(1)} =0 \,.
\ee
Equations \rf{21052020-man-16} tell us that the vertices $V^{(1)}$, $\Vb^{(1)}$ do not depend on the momenta $\beta_1$, $\beta_2$, $\beta_3$,
\be \label{21052020-man-17}
V^{(1)} = C \,, \qquad     \hspace{1cm} \Vb^{(1)} = \Cb \,.
\ee
We now should respect the requirement for $p_\smpth^-$ \rf{01032020-man-x01} to be $\beta$-analytic. From \rf{01032020-man-03}, we see that $p_\smpth^-$  is $\beta$-analytic iff $V_1$, $V_0$ are also $\beta$-analytic. Using \rf{01032020-man-11}, we get
\be \label{21052020-man-19}
V_0 = \half(V + \Vb)\,, \qquad V_1 = \frac{1}{2\kappa} (V - \Vb)\,.
\ee
Vertices $V$,$\Vb$ given in \rf{21052020-man-15} are polynomial in $\kappa$. From \rf{21052020-man-15},\rf{21052020-man-19}, and $\kappa^2$ \rf{01032020-man-x02}, we learn that $V_0$ is $\beta$-analytic provided $V_0$ is polynomial in $\kappa^2$. From \rf{21052020-man-19}, we see that this happens for $C=\Cb$. For such choice of $C$, $\Cb$, we see that $V_1$ \rf{21052020-man-19} is also $\beta$-analytic.

Procedure of the derivation of all remaining vertices is the same we have just described for vertex \rf{02032020-man-12}. Therefore to avoid the repetitions, below we present explicit form of transformations from the vertices $V$, $\Vb$ to vertices $V^{(1)}$, $\Vb^{(1)}$ which do not depend on $\beta_1$, $\beta_2$, $\beta_3$.

\noindent {\bf Cubic vertices \rf{02032020-man-19}, \rf{02032020-man-19-a}}.  The $\kappa$ \rf{01032020-man-x02} is chosen to be $\kappa = \Po_{\epsilon m}$. Using notation in \rf{02032020-man-20}, \rf{02032020-man-21}, we note the transformations and helpful relation,
\beq
\label{21052020-man-25} && V = \Po_{\epsilon m}^{\, \Sbf_\epsilon } \prod_{a=1,2,3} \beta_a^{-s_{\epsilon a}} V^{(1)}\,, \qquad \Vb = \Po_{\epsilon m}^{-\Sbf_\epsilon} \prod_{a=1,2,3} \beta_a^{s_{\epsilon a}} \Vb^{(1)}\,,
\\
&&  -  \So_\epsilon \Po_{\epsilon m} + \frac{\beta}{3} \Sbf_\epsilon   \PP_{\epsilon m} =  \beta \MM_{sm}\,, \qquad \MM_{s m} \equiv \sum_{a=1,2,3} \frac{s_a m_a}{\beta_a}\,.
\eeq

\noindent {\bf Cubic vertex \rf{02032020-man-27}}.
\be \label{21052020-man-27}
V =   L_{\kappa,1}^{s_1} L_{\kappa,2}^{s_2} V^{(1)}\,, \hspace{1cm}
\Vb =   L_{-\kappa,1}^{s_1} L_{-\kappa,2}^{s_2} \Vb^{(1)}\,.
\ee

\noindent {\bf Cubic vertices \rf{02032020-man-36}-\rf{02032020-man-39}}. For illustration, consider vertices \rf{02032020-man-36},\rf{02032020-man-37}. The $\kappa$ \rf{01032020-man-x02} is chosen to be $\kappa = m_1 \beta_3$. The transformations and equations \rf{01032020-man-x03}  take the respective forms
\beq
\label{21052020-man-29} && V = (\frac{\beta_3}{\beta_1})^{-s_1}  (\frac{\beta_3}{\beta_2})^{s_2} V^{(1)}\,,
\hspace{0.5cm} \Vb = (\frac{\beta_3}{\beta_1})^{s_1}  (\frac{\beta_3}{\beta_2})^{-s_2}  \Vb^{(1)}\,,
\\
\label{21052020-man-28} &&  \No_\beta V = (s_1 \beta_2 + s_2\beta_1) V\,,\hspace{1cm} \No_\beta \Vb = -(s_1 \beta_2 + s_2\beta_1) \Vb\,.
\eeq
Solution for $V$ and $\Vb$ turns out to be $\beta$-analytic without any constraints on $V^{(1)}$ and $\Vb^{(1)}$. The solution for vertices $V$, $\Vb$ leads to the respective two vertices \rf{02032020-man-36} and \rf{02032020-man-37}.

\noindent {\bf Cubic vertex \rf{02032020-man-45}}.
\be \label{21052020-man-30}
V =   L_{\kappa,3}^{s_3}  V^{(1)}\,, \hspace{1cm}
\Vb =   L_{-\kappa,3}^{s_3} \Vb^{(1)}\,.
\ee
\noindent {\bf Cubic vertex \rf{03032020-man-12}}. Helpful relation is given by \rf{21052020-man-02}. The transformations take the form
\be \label{21052020-man-45}
V =  \beta_1^{-\half} \beta_2^{-\half} L_{\kappa,1}^{s_1+\half} L_{\kappa,2}^{s_2+\half} L_{\kappa,3}^{s_3} V^{(1)}\,, \hspace{1cm}
\Vb =  \beta_1^{-\half} \beta_2^{-\half} L_{-\kappa,1}^{s_1+\half} L_{-\kappa,2}^{s_2+\half} L_{-\kappa,3}^{s_3} \Vb^{(1)}\,.
\ee
\noindent {\bf Cubic vertices \rf{03032020-man-19}, \rf{03032020-man-19-a}}. The $\kappa$ \rf{01032020-man-x02} is chosen to be $\kappa = \Po_{\epsilon m}$. The transformations take the form
\be \label{21052020-man-46}
V = \beta_1^{-\half} \beta_2^{-\half} \Po_{\epsilon m}^{\, \Sbf_\epsilon^\f } \prod_{a=1,2,3} \beta_a^{-s_{\epsilon a}^\f} V^{(1)}\,,  \hspace{0.5cm}
\Vb = \beta_1^{-\half} \beta_2^{-\half} \Po_{\epsilon m}^{-\Sbf_\epsilon^\f} \prod_{a=1,2,3} \beta_a^{s_{\epsilon a}^\f} \Vb^{(1)}\,.
\ee
\noindent {\bf Cubic vertex \rf{03032020-man-27}}. The helpful relation is obtained by setting $m_3=0$ in \rf{21052020-man-02}. The transformations take the form
\be \label{21052020-man-47}
V =   \beta_1^{-\half} \beta_2^{-\half} L_{\kappa,1}^{s_1+\half} L_{\kappa,2}^{s_2+\half} V^{(1)}\,, \hspace{1cm}
\Vb =   \beta_1^{-\half} \beta_2^{-\half} L_{-\kappa,1}^{s_1+\half} L_{-\kappa,2}^{s_2+\half} \Vb^{(1)}\,.
\ee
\noindent {\bf Cubic vertex \rf{03032020-man-35}}. The helpful relation  and the transformations take the form
\be \label{21052020-man-47-c}
V =   \beta_1^{-\half} \beta_2^{-\half} L_{\kappa,2}^{s_2+\half} L_{\kappa,3}^{s_3} V^{(1)}\,, \hspace{0.5cm}
\Vb =   \beta_1^{-\half} \beta_2^{-\half} L_{-\kappa,2}^{s_2+\half} L_{-\kappa,3}^{s_3} \Vb^{(1)}\,, \hspace{0.5cm} \frac{1}{\beta_1\beta_2}L_{\kappa,2}    =   - \frac{1}{2 m_2}  K_\kappa^2\,.
\ee
\noindent {\bf Cubic vertices \rf{03032020-man-43}-\rf{03032020-man-46}}. For illustration, consider vertices \rf{03032020-man-43},\rf{03032020-man-44}. The $\kappa$ \rf{01032020-man-x02} is chosen to be $\kappa = m_1 \beta_3$. The transformations and equations for $V$, $\Vb$ \rf{01032020-man-x03} take the respective forms
\beq
\label{21052020-man-49} && V = \frac{1}{\beta_2}(\frac{\beta_3}{\beta_1})^{-s_1}  (\frac{\beta_3}{\beta_2})^{s_2} V^{(1)}\,,
\hspace{0.5cm} \Vb = \frac{1}{\beta_1}(\frac{\beta_3}{\beta_1})^{s_1}  (\frac{\beta_3}{\beta_2})^{-s_2} \Vb^{(1)}\,,
\\
\label{21052020-man-48} && \No_\beta^E V = \big( (s_1 +\half)\beta_2 + (s_2+\half) \beta_1\big) V\,, \hspace{0.5cm} \No_\beta^E \Vb = -\big( (s_1+\half) \beta_2 + (s_2+\half)\beta_1 \big) \Vb\,,\qquad
\eeq
where $\No_\beta^E \equiv\No_\beta - \frac{1}{6}\betach_3$, while $\No_\beta$ is given in \rf{29022020-man-13}. Solution for $V$ and $\Vb$ turns out to be $\beta$-analytic without any constraints on $V^{(1)}$ and $\Vb^{(1)}$. The solution for vertices $V$, $\Vb$ leads to the respective two vertices  \rf{03032020-man-43} and \rf{03032020-man-44}.

\noindent {\bf Cubic vertices \rf{03032020-man-53}-\rf{03032020-man-56}}. For illustration, consider vertices \rf{03032020-man-53},\rf{03032020-man-54}.  The $\kappa$  \rf{01032020-man-x02} is chosen to be $\kappa = m_3 \beta_1$. The transformation and equations \rf{01032020-man-x03}  take the respective forms
\beq
\label{21052020-man-51} && V = \frac{1}{\beta_1}(\frac{\beta_1}{\beta_2})^{-s_2}  (\frac{\beta_1}{\beta_3})^{s_3} V^{(1)}\,, \hspace{0.5cm} \Vb = \frac{1}{\beta_2}(\frac{\beta_1}{\beta_2})^{s_2}  (\frac{\beta_1}{\beta_3})^{-s_3} \Vb^{(1)}\,,
\\
\label{21052020-man-50}  &&
\No_\beta^E V = \big( (s_2 +\half) \beta_3 + s_3\beta_2\big) V\,, \hspace{1cm}  \No_\beta^E \Vb = -\big( (s_2 +\half) \beta_3 + s_3\beta_2 \big) \Vb\,.
\eeq
Solution for $V$ and $\Vb$ turns out to be $\beta$-analytic without any restrictions on $V^{(1)}$ and $\Vb^{(1)}$. The solution for vertices $V$, $\Vb$ leads to the respective two vertices  \rf{03032020-man-53} and \rf{03032020-man-54}.

\noindent {\bf Cubic vertex \rf{03032020-man-63}}.
\be \label{21052020-man-52}
V =   \frac{1}{\kappa} L_{\kappa,3}^{s_3}  V^{(1)}\,, \hspace{1cm}
\Vb =  -\frac{1}{\kappa} L_{-\kappa,3}^{s_3} \Vb^{(1)}\,.
\ee

\noindent {\bf Cubic vertex \rf{03032020-man-70}}. The transformation and helpful relation take the form
\be
\label{21052020-man-53}   V =     \beta_1^{-\half}\beta_2^{-\half} L_{\kappa,1}^{s_1+\half}  V^{(1)}\,, \hspace{0.6cm}
\Vb =   \beta_1^{-\half}\beta_2^{-\half} L_{-\kappa,1}^{s_1+\half} \Vb^{(1)}\,, \hspace{0.5cm} \frac{1}{\beta_1\beta_2} L_{\kappa,1}   =   \frac{1}{2 m_1}  K_\kappa^2\,.
\ee

\noindent {\bf Interrelation between first-derivative and higher-derivative cubic vertices}. We now clarify our procedure of derivation of higher-derivative cubic vertex \rf{01032020-man-12} from first-derivative vertex \rf{01032020-man-x01}. This procedure is straightforward when the first-derivative vertex is polynomial in $\kappa^2$. Namely, taking into account definition of $\kappa^2$ \rf{01032020-man-x02}, we represent relation \rf{29022020-man-11} as
\be \label{21052020-man-65-01}
\kappa^2 =  - \Po^2 +  2\beta \Pbf^-  \,.
\ee
From \rf{21052020-man-65-01}, we see that, if first-derivative vertex $p_\smpth^-(\kappa)$ \rf{01032020-man-x01} is polynomial in $\kappa^2$, then the following  relation holds true
\be \label{21052020-man-65}
p_\smpth^-(\kappa)\bigr|_{\kappa^2 = - \Po^2 + 2\beta \Pbf^-} = p_\smpth^-(\irm \Po) + \Pbf^- f\,,
\ee
where $f$ is polynomial in $\Po$. The term $\Pbf^- f$ in \rf{21052020-man-65} can be removed by using field redefinitions (see \rf{01032020-man-01}). This implies that  higher-derivative vertex $V_\HDsm \equiv p_\smpth^-(\irm \Po)$ \rf{01032020-man-12} is obtainable from first-derivative vertex $p_\smpth^-(\kappa)$ \rf{01032020-man-x01} by using field redefinitions.

The substitution $\kappa=\irm \Po$ is straightforward when the first-derivative vertex is expressed in terms of quantities $L_{\kappa,a}$ \rf{02032020-man-13}. However some first-derivative vertices are not expressed in terms of $L_{\kappa,a}$ (see bose vertices \rf{02032020-man-19},\rf{02032020-man-19-a}, \rf{02032020-man-36}-\rf{02032020-man-39} and fermi-bose vertices \rf{03032020-man-19}, \rf{03032020-man-19-a}, \rf{03032020-man-43}-\rf{03032020-man-46}, \rf{03032020-man-53}-\rf{03032020-man-56}). To illustrate our method for these cases, we consider vertices  \rf{02032020-man-19}, \rf{02032020-man-36}.

\noindent {\bf First-derivative vertices \rf{02032020-man-19},\rf{02032020-man-19-a} and higher-derivative vertices \rf{02032020-man-23-b},\rf{02032020-man-24-b}}. For this case, the $\kappa$ \rf{01032020-man-x02} is chosen to be $\kappa = \Po_{\epsilon m}$. By using \rf{21052020-man-65-01}, we get then the relation
\be \label{21052020-man-66}
\Po_{\epsilon m}^2  =   - \Po^2   + 2 \beta \Pbf^- \,.
\ee
Upon substitution \rf{21052020-man-66} in cubic vertex, we can ignore $\Pbf^-$-terms because all contributions proportional to $\Pbf^-$ can be removed by field redefinitions. Thus, we have the equivalence relation $\Po_{\epsilon m}^2  \sim   - \Po^2$ for the cubic vertices. Using this equivalence relation, we get
\be \label{21052020-man-67}
\big( \irm \Po +  \Po_{\epsilon m} \big)^2    \sim  2\big( \irm \Po +  \Po_{\epsilon m} \big) \Po_{\epsilon m} \,,\qquad \big( \irm \Po -  \Po_{\epsilon m} \big)^2    \sim  - 2\big( \irm \Po -  \Po_{\epsilon m} \big) \Po_{\epsilon m} \,.
\ee
Using \rf{21052020-man-67}, we find that higher-derivative vertices \rf{02032020-man-23-b},\rf{02032020-man-24-b} amount to first-derivative vertices \rf{02032020-man-19},\rf{02032020-man-19-a}.

\noindent {\bf First-derivative vertex \rf{02032020-man-36} and higher-derivative vertex \rf{02032020-man-40}}. For masses \rf{02032020-man-35}, the $\kappa^2$ \rf{01032020-man-x02} takes the form $\kappa^2=m_1^2 \beta_3^2$. Relation \rf{21052020-man-65-01} implies then the following equivalence relation
$\Po^2 \sim - m_1^2 \beta_3^2$. Using this equivalence relation, we find the desired equivalence relation
\be \label{21052020-man-69}
\big( \frac{\irm \Po + m_2 \beta_3}{\beta_2} \big)^{s_2} \sim   2^{s_2-1} m_2^{s_2} \big(\frac{\beta_3}{\beta_2}\big)^{s_2} \big( 1+  \frac{\irm \Po}{m_2 \beta_3} )\,.
\ee
Using \rf{21052020-man-69} for first-derivative vertex \rf{02032020-man-36} gives higher-derivative vertex \rf{02032020-man-40}.

\noindent {\bf First-derivative vertex \rf{03032020-man-63} and higher-derivative vertex \rf{03032020-man-66}}. Finally we make comment on vertex \rf{03032020-man-63}. This vertex  depends on $L_{\kappa, 3}$ however is not polynomial in $\kappa$ ($\kappa$ appears in the denominator in \rf{03032020-man-64}). For this case we represent $p_\smpth^-$ as $p_\smpth^-(\kappa) = \frac{1}{\kappa^2} \tilde{p}_\smpth^-(\kappa)$, where $\tilde{p}_\smpth^-(\kappa)$ is polynomial in $\kappa^2$. Then higher-derivative vertex \rf{03032020-man-66} is obtained simply as $p_\smpth^- = \frac{1}{\kappa^2} \tilde{p}_\smpth^-(\irm \Po)$.

%%%%%%%%%%%%%%%%%%%%%%%%%%%%%%%%%%%%%%%%%%%%%%%%%%%%%%%%%%%%%%%%%%%%%%%%%%%%%%%%%%%%%%%%%%%%%%%%%%%%%%%
%%%%%%%%%%%%%%%%%%%%%%%%%%%%%%%%%%%%%%%%%%%%%%%%%%%%%%%%%%%%%%%%%%%%%%%%%%%%%%%%%%%%%%%%%%%%%%%%%%%%%%%
\appendix{ \large Dimensional reduction from massless fields in $R^{3,1}$ to massive fields in $R^{2,1}$ }
%%%%%%%%%%%%%%%%%%%%%%%%%%%%%%%%%%%%%%%%%%%%%%%%%%%%%%%%%%%%%%%%%%%%%%%%%
%%%%%%%%%%%%%%%%%%%%%%%%%%%%%%%%%%%%%%%%%%%%%%%%%%%%%%%%%%%%%%%%%%%%%%%%%

We now outline how our bose vertices \rf{02032020-man-23-b},\rf{02032020-man-24-b} can be obtained via dimensional reduction from vertices for massless fields in Ref.\cite{Bengtsson:1986kh}. Generalization of our consideration to the fermi-bose vertices is straightforward. In helicity basis, bosonic spin-$s$, $s\geq 1$, massless field in $R^{3,1}$ is described by two complex-valued fields $\phi_{\pm s}(p^1,p^2,\beta)$,  $\phi_s(p^1,p^2,\beta)^\dagger = \phi_{-s}(-p^1,-p^2,-\beta)$, where $p^1$, $p^2$, $\beta$ are momenta and dependence on light-cone time $x^+$ is implicit.
To proceed with dimensional reduction we make Fourier transform with respect to the momentum $p^2$ and introduce fields $\phi_{\pm s}(p^1,x^2,\beta)$.%
\footnote{ Fourier transform and its inverse are chosen to be as follows $f(x)=\int\frac{dp}{\sqrt{2\pi}} e^{\irm px} \tilde{f}(p)$, $\tilde{f}(p)=\int\frac{dx}{\sqrt{2\pi}} e^{-\irm px} f(x)$.
}
Now making compactification on a circle $x^2 \rightarrow \varphi$, $\varphi\in (0,2\pi)$, we get fields $\phi_{\pm s}(p^1,\varphi,\beta)$. For such fields, we use the following Fourier expansion into infinite set of fields in $R^{2,1}$,
\be  \label{30052020-man-01}
\phi_\lambda(p^1,\varphi,\beta) = \sum_{n=-\infty}^{\infty} e^{\irm \epsilon m_n \varphi } \phi_{m_n,\lambda}(p^1,\beta)\,, \qquad m_n\equiv n\,,\qquad \lambda=\epsilon s\,, \qquad \epsilon =\pm 1\,,
\ee
where  $\phi_{m_n,\pm s}$ are two complex-valued fields. Hermicity conditions 
for fields \rf{30052020-man-01} take the form 
\be \label{30052020-man-01-a}
\phi_s(p^1,\varphi,\beta)^\dagger = \phi_{-s}(-p^1,\varphi,-\beta)\,, \qquad \phi_{m_n,s}(p^1,\beta)^\dagger = \phi_{m_n,-s}(-p^1,-\beta)\,.
\ee
The second relation in \rf{30052020-man-01-a} implies that two complex-valued fields $\phi_{m_n,\pm s}(p^1,\beta)$ can be represented in terms of two fields (which are real-valued in $x$-space)
\be   \label{30052020-man-02}
\phi_{m_n,s}  = \phi_{m_n,s}^\smone + \irm \phi_{m_n,s}^\smtwo\,, \hspace{1cm}
\phi_{m_n,-s}  = \phi_{m_n,s}^\smone - \irm \phi_{m_n,s}^\smtwo\,,
\ee
where two fields $\phi_{m_n,s}^\smone$, $\phi_{m_n,s}^\smtwo$ satisfy the hermicity condition as in \rf{27022020-man-12}. Now we show that $\phi_{m_n,s}^\smone$, $\phi_{m_n,s}^\smtwo$ can be identified with the massive fields in \rf{27022020-man-07}. To this end we prove that, under action of the Poinvcar\'e algebra $iso(2,1)$, the fields $\phi_{m_n,s}^\smone$, $\phi_{m_n,s}^\smtwo$ transform in the same way as fields \rf{27022020-man-07}. With the exception of $J^{-1}$-transformations, matching all transformations in \rf{27022020-man-13}-\rf{27022020-man-15} is obvious. To match $J^{-1}$-transformations we note that $J^{-1}$-transformations of fields $\phi_\lambda$ \rf{30052020-man-01} are realized by the following differential operator:%
\footnote{ To get \rf{30052020-man-03}, we use $J^{-1} =   (J^{-\Rsm} + J^{-\Lsm})/\sqrt{2}$ and $J^{-\Rsm}$, $J^{-\Lsm}$ given in (2.26), (2.27) in Ref.\cite{Metsaev:2019dqt}.}
\be   \label{30052020-man-03}
J^{-1} =- \partial_\beta p^1 + \partial_{p^1} p^- +   \frac{\irm\lambda p^2}{\beta} + \frac{p^1}{2\beta} e_\lambda \,, \hspace{0.5cm} p^- = -\frac{p^1p^1 + p^2p^2}{2\beta}\,, \hspace{0.5cm} p^2 = - \irm \partial_\varphi\,,
\ee
where $e_\lambda$ is relevant only for half-integer fields. Acting with $J^{-1}$ on $\phi_\lambda$ \rf{30052020-man-01} and using $\lambda=\epsilon s$, $p\equiv p^1$, we find that action of $J^{-1}$ on fields $\phi_{m_n,s}^\smone$, $\phi_{m_n,s}^\smtwo$ \rf{30052020-man-02} is realized by the differential operator
\be \label{30052020-man-04}
J^{-1} =- \partial_\beta p + \partial_{p} p^- +   \frac{\irm s m_n}{\beta} + \frac{p}{2\beta} e_\lambda \,, \hspace{0.5cm} p^- = -\frac{pp + m_n^2}{2\beta}\,.
\ee
Comparing \rf{30052020-man-04} with \rf{27022020-man-15}, we see that each field $\phi_{m_n,s}^\smone$ $\phi_{m_n,s}^\smtwo$ is indeed realized as spin-$s$ and mass $m_n$ field, while a basis of the fields $\phi_{m_n,\pm s}$ \rf{30052020-man-02} can be interpreted as some kind of helicity basis for massive fields in $R^{2,1}$.  Relation \rf{30052020-man-01} tells us that massless spin-$s$ field in $R^{3,1}$ described by the fields $\phi_{\pm s}$ is indeed amount to infinite set of Kaluza-Klein massive spin-$s$ fields $\phi_{m_n,s}^\smone$, $\phi_{m_n,s}^\smtwo$ in $R^{2,1}$ with the spectrum of masses $m_n=n$, $n=\pm1,\ldots, \pm\infty$ (plus two massless fields in $R^{2,1}$). For $\lambda=0$ in \rf{30052020-man-01}, we can use a  scalar field $\phi_0$, which is real valued in $x$-space, and fix $\epsilon = 1$.

We now turn to cubic vertices. All that is required is to plug \rf{30052020-man-01} into cubic vertices for massless fields in $R^{3,1}$ obtained in Ref.\cite{Bengtsson:1986kh}. Let us present those vertices by using the conventions in Ref.\cite{Metsaev:2019dqt}. In terms of the momentum-space fields, cubic vertex for three massless fields having helicities $\lambda_1$, $\lambda_2$, $\lambda_3$ takes the form
\be \label{30052020-man-05}
P_{\smpth;\lambda_1,\lambda_2,\lambda_3}^- = \int d\Gamma_\smpth \phi_{\lambda_1}^\dagger(\vec{p}_1) \phi_{\lambda_2}^\dagger(\vec{p}_2) \phi_{\lambda_3}^\dagger(\vec{p}_3) C^{\lambda_1\lambda_2\lambda_3}
\frac{(\Po^\Lsm)^{\lambda_1+\lambda_2+\lambda_3}}{\beta_1^{\lambda_1}\beta_2^{\lambda_2} \beta_3^{\lambda_3} } + h.c., \hspace{0.5cm} \Po^\Lsm = \frac{1}{\sqrt{2}}(\Po^1 - \irm \Po^2)\,,
\ee
where coupling constants $C^{\lambda_1\lambda_2\lambda_3}$ are complex-valued in general. The $\Po^1$, $\Po^2$ are defined as in \rf{29022020-man-03}, while an integration measure is given by $d\Gamma_\smpth = (2\pi)^3\delta^3(\vec{p}_1+\vec{p}_2+\vec{p}_3)\prod_{a=1,2,3}d^3\vec{p}_a/(2\pi)^{3/2}$, $\vec{p}_a=p_a^1,p_a^2,\beta_a$.
To proceed we should transform fields in \rf{30052020-man-05} from coordinates $p_a^1,p_a^2,\beta_a$ to coordinates $p_a^1,\varphi_a,\beta_a$. Doing so, we get fields as in \rf{30052020-man-01}. Using fields \rf{30052020-man-01} in \rf{30052020-man-05} and integrating out angle variables $\varphi_1$, $\varphi_2$, $\varphi_3$, we get
\beq
\label{30052020-man-06} &&  \hspace{-0.7cm} P_{\smpth;\lambda_1,\lambda_2,\lambda_3}^- = \sum_{n_1,n_2,n_3 = -\infty}^\infty \delta_{\Pbf_{\epsilon m},0} \int d\Gamma_\smpth p_{ m_{n_1},\lambda_1; m_{n_2},\lambda_2; m_{n_3},\lambda_3 }^- \prod_{a=1,2,3}\phi_{m_{n_a},\lambda_a}(\vec{p}_a) +h.c.\,,\qquad
\\
\label{30052020-man-07} && \hspace{-0.7cm}  p_{ m_{n_1},\lambda_1; m_{n_2},\lambda_2; m_{n_3},\lambda_3 }^- =  k^{\lambda_1\lambda_2\lambda_3} C^{\lambda_1\lambda_2\lambda_3}
\frac{\big(\irm \Po + \Po_{\epsilon m}\big)^{\lambda_1+\lambda_2+\lambda_3}}{\beta_1^{\lambda_1}\beta_2^{\lambda_2} \beta_3^{\lambda_3} }\,, \qquad \lambda_a = \epsilon_a s_a\,, \qquad
\eeq
$m_{n_a}=n_a$, $ k^{\lambda_1\lambda_2\lambda_3} = 2\pi (\irm \sqrt{2})^{-\lambda_1-\lambda_2-\lambda_3}$, where $\delta_{kn} = 1$ for $k=n$ and $\delta_{kn} = 0$ for $k\ne n$. The $d\Gamma_\smpth$ \rf{30052020-man-06} is obtained from \rf{28022020-man-04} by equating $n=3$. In \rf{30052020-man-06}, \rf{30052020-man-07}, we use $\vec{p_a} = p_a,\beta_a$ and $p_a=p_a^1$, $\Po\equiv \Po^1$. The $\Pbf_{\epsilon m}$ and $\Po_{\epsilon m}$ are obtained from  \rf{02032020-man-02}, \rf{02032020-man-21} by the substitution for masses $m_a \rightarrow  n_a$. In view of $\delta_{\Pbf_{\epsilon m},0}$ in \rf{30052020-man-06}, we note the appearance of the condition $\Pbf_{\epsilon m}=0$ for Kaluza-Klein masses.%
\footnote{ For $\epsilon_1=\epsilon_2=\epsilon_3$, the condition $\Pbf_{\epsilon m}=0$ takes the form $m_1+m_2+m_3=0$. The condition $m_1+m_2+m_2=0$  have also been encountered in Ref.\cite{Fotopoulos:2009iw} when studying totally symmetric fields in arbitrary dimensions. In our study all conditions $\Pbf_{\epsilon m}=0$ are realized. We cordially thank to M.Tsulaia for drawing our attention to Ref.\cite{Fotopoulos:2009iw}.
}

In Refs.\cite{Metsaev:1991mt,Metsaev:1991nb}, we studied massless higher-spin theory with the following cubic interactions:
\be
P_\smpth^- = \sum_{\lambda_1,\lambda_2,\lambda_3 = - \infty}^\infty P_{\smpth;\lambda_1,\lambda_2,\lambda_3}^-\,,
\ee
where $P_{\smpth;\lambda_1,\lambda_2,\lambda_3}^-$ is given in \rf{30052020-man-05}. Some solution for coupling constants $C^{\lambda_1\lambda_2\lambda_3}$ in Refs.\cite{Metsaev:1991mt,Metsaev:1991nb} was found. Using such solution and ignoring hermitian conjugated part in \rf{30052020-man-05}, the massless higher-spin chiral theory was proposed in Ref.\cite{Ponomarev:2016lrm}. By using the solution for coupling constants  $C^{\lambda_1\lambda_2\lambda_3}$ in Ref.\cite{Metsaev:1991mt,Metsaev:1991nb}, it seems then naturally to suggest massive higher-spin chiral theory by using \rf{30052020-man-06},\rf{30052020-man-07}, and removing hermitian conjugated part in  \rf{30052020-man-06}. Finally, we note that massive higher-spin theory with full (hermitian) Hamiltonian \rf{30052020-man-06},\rf{30052020-man-07} is also worth studying.

%%%%%%%%%%%%%%%%%%%%%%%%%%%%%%%%%%%%%%%%%%%%%%%%%%%%%%%%%%%%%%%%%%%%%%%%%%%%%%%%
%%%%%%%%%%%%%%%%%%%%%%%%%%%%%%%%%%%%%%%%%%%%%%%%%%%%%%%%%%%%%%%%%%%%%%%%%%%%%%%%

\end{document}